\UseRawInputEncoding
\documentclass[11pt]{article}
\usepackage{amsmath,amssymb,amsthm,amsxtra,overpic,bbm,bm,epsfig,ulem,color,multirow}
\usepackage{braket}

\begin{document}

\begin{center}
{\Large A combination of the neutrino trimaximal mixings \\
and $\mu$-$\tau$ reflection symmetry in the type-I seesaw model}
\end{center}

\vspace{0.05cm}

\begin{center}
{\bf Zhen-hua Zhao\footnote{zhaozhenhua@lnnu.edu.cn}} \\
{ Department of Physics, Liaoning Normal University, Dalian 116029, China }
\end{center}

\vspace{0.2cm}

\begin{abstract}
In this paper, we make an attempt to combine the neutrino trimaximal (TM1 and TM2) mixings and $\mu$-$\tau$ reflection symmetry in the type-I seesaw model. Such a scenario is highly restrictive and predictive: in addition to three right-handed neutrino masses, there are only five real parameters; all the lepton flavor mixing parameters except for $\theta^{}_{13}$ will be predicted. The relations between the model parameters and the measurable neutrino parameters will be derived.
The implications of this scenario for leptogenesis will be explored in detail. A further reduction of this scenario to the more restrictive and predictive minimal seesaw model with only two right-handed neutrinos will also be considered. In addition, we will also discuss a possible approach to get the desired mass matrices and study the compatibility of the trimaximal $\mu$-$\tau$ reflection symmetry with texture zeros.
\end{abstract}

\newpage

\section{Introduction}

As we know, the observation of neutrino oscillations tells us that neutrinos are massive and the lepton flavors are mixed \cite{xing}. On the one hand, the type-I seesaw model \cite{seesaw} is the most promising theory for the neutrino mass generation: three right-handed neutrino fields $N^{}_I$ (for $I=1, 2, 3$) are introduced; they not only have the Yukawa couplings $(Y^{}_\nu)^{}_{\alpha I}$ with three left-handed neutrino fields $\nu^{}_\alpha$ (for $\alpha = e, \mu, \tau$), which lead to the Dirac neutrino mass matrix $M^{}_{\rm D} = Y^{}_\nu v$ (with $v = 174$ GeV being the vacuum expectation value of the Higgs field $H$), but also have a Majorana mass matrix $M^{}_{\rm R}$ on their own. After integrating out the heavy right-handed neutrino fields, one will obtain an effective Majorana mass matrix as $M^{}_{\nu} \simeq - M^{}_{\rm D} M^{-1}_{\rm R} M^{T}_{\rm D}$ for three light neutrinos.
Without loss of generality, we will work in the basis of $M^{}_{\rm R}$ being diagonal $D^{}_{\rm R} = {\rm diag}(M^{}_1, M^{}_2, M^{}_3)$ with $M^{}_I$ being three right-handed neutrino masses.

On the other hand, in the basis where the mass eigenstates of three charged leptons are identical with their flavor eigenstates, the lepton flavor mixing matrix $U$ arises as the unitary matrix for diagonalizing $M^{}_\nu$: $U^\dagger M^{}_\nu U^* =  D^{}_\nu =  {\rm diag}(m^{}_1, m^{}_2, m^{}_3) $ with $m^{}_i$ being three light neutrino masses.
In the standard parametrization, $U$ is expressed in terms of three lepton flavor mixing angles $\theta^{}_{ij}$ (for $ij=12, 13, 23$), the Dirac CP phase $\delta$ and two Majorana CP phases $\rho$ and $\sigma$:
\begin{eqnarray}
U  = \left( \begin{matrix}
c^{}_{12} c^{}_{13} & s^{}_{12} c^{}_{13} & s^{}_{13} e^{-{\rm i} \delta} \cr
-s^{}_{12} c^{}_{23} - c^{}_{12} s^{}_{23} s^{}_{13} e^{{\rm i} \delta}
& c^{}_{12} c^{}_{23} - s^{}_{12} s^{}_{23} s^{}_{13} e^{{\rm i} \delta}  & s^{}_{23} c^{}_{13} \cr
s^{}_{12} s^{}_{23} - c^{}_{12} c^{}_{23} s^{}_{13} e^{{\rm i} \delta}
& -c^{}_{12} s^{}_{23} - s^{}_{12} c^{}_{23} s^{}_{13} e^{{\rm i} \delta} & c^{}_{23}c^{}_{13}
\end{matrix} \right) \left( \begin{matrix}
e^{{\rm i}\rho} &  & \cr
& e^{{\rm i}\sigma}  & \cr
&  & 1
\end{matrix} \right) \;,
\label{1}
\end{eqnarray}
where $c^{}_{ij} = \cos \theta^{}_{ij}$ and $s^{}_{ij} = \sin \theta^{}_{ij}$ have been used for notational simplicity.

Now, thanks to the various neutrino oscillation experiments, the neutrino mass squared differences $\Delta m^2_{ij} \equiv m^2_i - m^2_j$ and lepton flavor mixing angles have been measured with a good degree of accuracy. And there has also been a preliminary result for $\delta$. The global-fit results for these parameters are presented in Table~\ref{tab1} \cite{global,global2}. However, the sign of $\Delta m^2_{31}$ remains unknown, allowing for two possible neutrino mass orderings: the normal ordering (NO) $m^{}_1 < m^{}_2 < m^{}_3$ and inverted ordering (IO) $m^{}_3 < m^{}_1 < m^{}_2$. And there has not been any lower constraint on the lightest neutrino mass, nor any constraint on the Majorana CP phases, whose values can only be inferred from some non-oscillatory processes such as the neutrinoless double beta decays \cite{0nbb}.

From Table~\ref{tab1} we see that $\theta^{}_{12}$ and $\theta^{}_{23}$ are around some special values: $\sin^2 \theta^{}_{12} \sim 1/3$ and $\sin^2 \theta^{}_{23} \sim 1/2$. Historically, $\theta^{}_{13}$ was widely expected to be vanishingly small before its value was precisely known. At that time, the tribimaximal (TBM) mixing \cite{TB}
\begin{eqnarray}
U^{}_{\rm TBM}= \displaystyle \frac{1}{\sqrt 6} \left( \begin{array}{ccc} \vspace{0.15cm}
2 & \sqrt{2} & 0 \cr \vspace{0.15cm}
1 & - \sqrt{2}  & -\sqrt{3}  \cr
1 & - \sqrt{2}  & \sqrt{3} \cr
\end{array} \right)  \;,
\label{2}
\end{eqnarray}
was very popular for its simple form and interesting predictions: $\sin^2 \theta^{}_{12} = 1/3$, $\sin^2 \theta^{}_{23} = 1/2$ (i.e., $\theta^{}_{23} = \pi/4$) and $\theta^{}_{13} =0$. Unfortunately, the relative largeness of $\theta^{}_{13}$ forces us to forsake or modify such an attractive mixing pattern.
An economical way out is to keep its first or second column unchanged while modifying the other two columns, thus giving the first or second trimaximal (TM1 or TM2) mixing \cite{TM}
\begin{eqnarray}
U^{}_{\rm TM1}=  \displaystyle \frac{1}{\sqrt 6} \left( \begin{array}{ccc} \vspace{0.15cm}
2 & \cdot & \cdot \cr \vspace{0.15cm}
1 & \cdot & \cdot \cr
1 & \cdot & \cdot \cr
\end{array} \right)  \;, \hspace{1cm}
U^{}_{\rm TM2}=  \displaystyle \frac{1}{\sqrt 3} \left( \begin{array}{ccc} \vspace{0.15cm}
\cdot & 1 & \cdot \cr \vspace{0.15cm}
\cdot & -1 &  \cdot \cr
\cdot &  -1 & \cdot \cr
\end{array} \right)  \;,
\label{3}
\end{eqnarray}
which basically preserve the TBM prediction for $\theta^{}_{12}$:
\begin{eqnarray}
{\rm TM1}: \hspace{0.5cm} s^{2}_{12} =  \frac{ 1}{3} -  \frac{2 s^{2}_{13}}{3 - 3s^{2}_{13}} \simeq 0.318 \;; \hspace{1cm} {\rm TM2}: \hspace{0.5cm}  s^{2}_{12} = \frac{ 1}{3} + \frac{s^{2}_{13}}{3 - 3s^{2}_{13}} \simeq 0.341 \;.
\label{4}
\end{eqnarray}

On the other hand, it is known that the $\mu$-$\tau$ symmetry (which requires the neutrino mass matrix to keep invariant under $\nu^{}_{\mu} \leftrightarrow \nu^{}_\tau$ \cite{mutau1, mutau2}) can naturally accommodate $\theta^{}_{23} = \pi/4$ and $\theta^{}_{13} =0$. After the observation of a relatively large $\theta^{}_{13}$ and a preliminary hint for $\delta \sim - \pi/2$ \cite{T2K}, its variant --- the $\mu$-$\tau$ reflection symmetry \cite{mu-tauR, mutau2} has attracted particular attention, which gives the following interesting predictions for the lepton flavor mixing parameters
\begin{eqnarray}
\theta^{}_{23} = \frac{\pi}{4} \;, \hspace{1cm} \delta = \pm \frac{\pi}{2} \;,
\hspace{1cm} \rho, \sigma = 0 \ {\rm or} \ \frac{\pi}{2} \;.
\label{5}
\end{eqnarray}
This symmetry requires the neutrino mass matrix to keep invariant under the following transformations of three left-handed neutrino fields
\begin{eqnarray}
\nu^{}_{e} \leftrightarrow \nu^{c}_e \;, \hspace{1cm} \nu^{}_{\mu} \leftrightarrow \nu^{c}_{\tau} \;,
\hspace{1cm} \nu^{}_{\tau} \leftrightarrow \nu^{c}_{\mu} \;,
\label{6}
\end{eqnarray}
where the superscript $c$ denotes the charge conjugation of relevant fields.

Motivated by the above facts, Rodejohann and Xu have considered a combination of the TM1 symmetry (i.e., the flavor symmetry responsible for the TM1 mixing) and $\mu$-$\tau$ reflection symmetry in Ref.~\cite{TMmutau}. Such a scenario is highly restrictive and predictive: all the lepton flavor mixing parameters except for $\theta^{}_{13}$ will be predicted (see Eqs.~(\ref{4}, \ref{5})). In this paper, we perform a further study of this interesting scenario from the following several aspects:
\begin{itemize}
\item while the study in Ref.~\cite{TMmutau} is performed at the Majorana neutrino mass matrix (i.e., $M^{}_\nu$) level, we will formulate the combination of the trimaximal mixings and $\mu$-$\tau$ reflection symmetry in the complete type-I seesaw model and derive the relations between the model parameters and the measurable neutrino parameters;
\item while the study in Ref.~\cite{TMmutau} focuses on the combination of the TM1 mixing and $\mu$-$\tau$ reflection symmetry, we will also study its TM2 counterpart;
\item the implications of this scenario for leptogenesis will be explored in detail;
\item a further reduction of this scenario to the more restrictive and predictive minimal seesaw model with only two right-handed neutrinos will be considered;
\item in addition, a possible approach to get the desired mass matrices will be discussed;
\item and the compatibility of the trimaximal $\mu$-$\tau$ reflection symmetry with texture zeros will also be studied.
\end{itemize}
The rest part of this paper is organized as follows: in the next section, we will formulate the combination of the trimaximal mixings and $\mu$-$\tau$ reflection symmetry in the seesaw model and derive the relations between the model parameters and the measurable neutrino parameters. In this section, we will also discuss a possible approach to get the desired mass matrices and study the compatibility of the trimaximal $\mu$-$\tau$ reflection symmetry with texture zeros.
In section~3, we explore the implications of this scenario for leptogenesis. In section~4, a further reduction of this scenario to the minimal seesaw model will be considered.

\begin{table}\centering
  \begin{footnotesize}
    \begin{tabular}{c|cc|cc}
     \hline\hline
      & \multicolumn{2}{c|}{Normal Ordering}
      & \multicolumn{2}{c}{Inverted Ordering }
      \\
      \cline{2-5}
      & bf $\pm 1\sigma$ & $3\sigma$ range
      & bf $\pm 1\sigma$ & $3\sigma$ range
      \\
      \cline{1-5}
      \rule{0pt}{4mm}\ignorespaces
       $\sin^2\theta^{}_{12}$
      & $0.318_{-0.016}^{+0.016}$ & $0.271 \to 0.370$
      & $0.318_{-0.016}^{+0.016}$ & $0.271 \to 0.370$
      \\[1mm]
       $\sin^2\theta^{}_{23}$
      & $0.566_{-0.022}^{+0.016}$ & $0.441 \to 0.609$
      & $0.566_{-0.023}^{+0.018}$ & $0.446 \to 0.609$
      \\[1mm]
       $\sin^2\theta^{}_{13}$
      & $0.02225_{-0.00078}^{+0.00055}$ & $0.02015 \to 0.02417$
      & $0.02250_{-0.00076}^{+0.00056}$ & $0.02039 \to 0.02441$
      \\[1mm]
       $\delta/\pi$
      & $1.20_{-0.14}^{+0.23}$ & $0.80 \to 2.00$
      & $1.54_{-0.13}^{+0.13}$ & $1.14 \to 1.90$
      \\[3mm]
       $\Delta m^2_{21}/(10^{-5}~{\rm eV}^2)$
      & $7.50_{-0.20}^{+0.22}$ & $6.94 \to 8.14$
      & $7.50_{-0.20}^{+0.22}$ & $6.94 \to 8.14$
      \\[3mm]
       $|\Delta m^2_{31}|/(10^{-3}~{\rm eV}^2)$
      & $2.56_{-0.04}^{+0.03}$ & $2.46 \to 2.65$
      & $2.46_{-0.03}^{+0.03}$ & $2.37 \to 2.55$
      \\[2mm]
      \hline\hline
    \end{tabular}
  \end{footnotesize}
  \caption{The best-fit values, 1$\sigma$ errors and 3$\sigma$ ranges of six neutrino
oscillation parameters extracted from a global analysis of the existing
neutrino oscillation data \cite{global}. }
\label{tab1}
\end{table}

\section{Trimaximal $\mu$-$\tau$ reflection symmetry in the seesaw model}

In this section, we formulate the combination of the trimaximal mixings and $\mu$-$\tau$ reflection symmetry in the seesaw model and derive the relations between the model parameters and the measurable neutrino parameters.

\subsection{TM1 $\mu$-$\tau$ reflection symmetry in the seesaw model}

In this subsection, we study the combination of the TM1 mixing and $\mu$-$\tau$ reflection symmetry.
In the seesaw model, the TM1 mixing can be naturally realized by taking one column of $M^{}_{\rm D}$ to be proportional to the first column of $U^{}_{\rm TBM}$ and the other two columns to be orthogonal to it \cite{TMss}:
\begin{eqnarray}
M^{}_{\rm D}= \left( \begin{array}{ccc}
2 a \sqrt{M^{}_1} & b \sqrt{M^{}_2} & d \sqrt{M^{}_3} \cr
a \sqrt{M^{}_1} & (-b-c) \sqrt{M^{}_2} & (-d-e)  \sqrt{M^{}_3} \cr
a \sqrt{M^{}_1} & (-b+c) \sqrt{M^{}_2} & (-d+e) \sqrt{M^{}_3} \cr
\end{array} \right) \;,
\label{2.1.1}
\end{eqnarray}
where $a, b, c, d$ and $e$ are generally complex. One can directly verify that the resulting $M^{}_\nu$ from such an $M^{}_{\rm D}$ does obey the TM1 symmetry:
\begin{eqnarray}
M^{}_\nu = R^{}_{\rm TM1} M^{}_\nu R^{}_{\rm TM1}
\hspace{0.5cm} {\rm with} \hspace{0.5cm}
R^{}_{\rm TM1}= - \frac{1}{3} \left( \begin{array}{ccc}
1 & 2 & 2 \cr
2 & -2 & 1 \cr
2 & 1 & -2 \cr
\end{array} \right)  \;.
\label{2.1.2}
\end{eqnarray}
On the other hand, the $\mu$-$\tau$ reflection symmetry can be naturally realized by taking $M^{}_{\rm D}$ to have a form as \cite{mutau2}
\begin{eqnarray}
M^{}_{\rm D} = \left( \begin{array}{ccc}
a \sqrt{M^{}_1} & c \sqrt{M^{}_2} & e \sqrt{M^{}_3} \cr
b \sqrt{M^{}_1} & d \sqrt{M^{}_2} & f \sqrt{M^{}_3}  \cr
b^* \sqrt{M^{}_1} & d^* \sqrt{M^{}_2} & f^* \sqrt{M^{}_3} \cr
\end{array} \right) P^{}_N  \;,
\label{2.1.3}
\end{eqnarray}
where $a, c$ and $e$ are real while $b, d$ and $f$ are generally complex, and $P^{}_N ={\rm diag}(\sqrt{\eta^{}_1}, \sqrt{\eta^{}_2}, 1)$ (for $\eta^{}_1, \eta^{}_2 = \pm 1$).
One can also directly verify that the resulting $M^{}_\nu$ from such an $M^{}_{\rm D}$ does obey the $\mu$-$\tau$ reflection symmetry:
\begin{eqnarray}
M^{}_\nu = R^{}_{\mu\tau} M^{*}_\nu R^{}_{\mu\tau} \hspace{0.5cm} {\rm with} \hspace{0.5cm}  R^{}_{\mu\tau}= \left( \begin{array}{ccc}
1 & 0 & 0 \cr
0 & 0 & 1 \cr
0 & 1 & 0 \cr
\end{array} \right)  \;.
\label{2.1.4}
\end{eqnarray}

Then, it is easy to see that, in order for $M^{}_{\rm D}$ to be consistent with the patterns described by Eqs.~(\ref{2.1.1}, \ref{2.1.3}) simultaneously, it should take a form as
\begin{eqnarray}
&& M^{}_{\rm D}= \left( \begin{array}{ccc}
2 a \sqrt{M^{}_1} & b \sqrt{M^{}_2} & d \sqrt{M^{}_3}  \cr
a \sqrt{M^{}_1} &  (-b- {\rm i} c) \sqrt{M^{}_2} & (-d - {\rm i} e)  \sqrt{M^{}_3} \cr
a \sqrt{M^{}_1} & (- b+ {\rm i} c) \sqrt{M^{}_2} & (-d + {\rm i} e) \sqrt{M^{}_3} \cr
\end{array} \right) P^{}_N \;,
\label{2.1.5}
\end{eqnarray}
with $a, b, c, d$ and $e$ being real.
The resulting $M^{}_\nu$ from this $M^{}_{\rm D}$ is given by
\begin{eqnarray}
&& M^{}_\nu = - \left( \begin{array}{ccc}
4 \eta^{}_1 a^2+ T^{}_1 & 2 \eta^{}_1 a^2 - T^{}_1 - {\rm i}T^{}_3 & 2 \eta^{}_1 a^2 - T^{}_1 + {\rm i}T^{}_3 \cr
2 \eta^{}_1 a^2 - T^{}_1 - {\rm i}T^{}_3 & \eta^{}_1 a^2 + T^{}_1 - T^{}_2 + 2  {\rm i} T^{}_3 & \eta^{}_1 a^2 + T^{}_1 + T^{}_2  \cr
2 \eta^{}_1 a^2 - T^{}_1 + {\rm i}T^{}_3 & \eta^{}_1 a^2 + T^{}_1 + T^{}_2  & \eta^{}_1 a^2 + T^{}_1 - T^{}_2 - 2  {\rm i} T^{}_3 \cr
\end{array} \right) \;,
\label{2.1.6}
\end{eqnarray}
with
\begin{eqnarray}
T^{}_1 = \eta^{}_2 b^2+ d^2 \;, \hspace{1cm} T^{}_2 =\eta^{}_2 c^2+ e^2 \;, \hspace{1cm}  T^{}_3 =\eta^{}_2 bc+ de  \; .
\label{2.1.7}
\end{eqnarray}
One can directly verify that such an $M^{}_\nu$ does obey the TM1 and $\mu$-$\tau$ reflection symmetries simultaneously. This $M^{}_\nu$ can be diagonalized by the following unitary matrix
\begin{eqnarray}
U^{}_0= \displaystyle \frac{1}{\sqrt 6} \left( \begin{array}{ccc} \vspace{0.15cm}
2 & \sqrt{2} & 0 \cr \vspace{0.15cm}
1 & - \sqrt{2}  & -\sqrt{3}  \cr
1 & - \sqrt{2}  & \sqrt{3} \cr
\end{array} \right)  \left( \begin{array}{ccc}
1 & 0 & 0 \cr
0 & 1  & 0  \cr
0 & 0  & {\rm i} \cr
\end{array} \right) \left( \begin{array}{ccc}
1 & 0 & 0 \cr
0 & \cos \theta  & \sin \theta  \cr
0 & -\sin \theta  & \cos \theta \cr
\end{array} \right) \; ,
\label{2.1.8}
\end{eqnarray}
with
\begin{eqnarray}
\tan 2 \theta = \frac{2 \sqrt{6} T^{}_3 }{2T^{}_2 - 3 T^{}_1} \;,
\label{2.1.9}
\end{eqnarray}
giving the neutrino masses as
\begin{eqnarray}
m^{\prime}_1= - 6 \eta^{}_1 a^2 \;, \hspace{1cm} m^{\prime}_{2, 3} = \frac{1}{2} \left[  -(3T^{}_1 + 2 T^{}_2) \pm \sqrt{ ( 3T^{}_1 + 2 T^{}_2)^2 + 24 (T^{2}_3 -T^{}_1 T^{}_2 )  }  \right]   \;.
\label{2.1.10}
\end{eqnarray}
With the help of Eqs.~(\ref{2.1.9}, \ref{2.1.10}) and $\sin^2 \theta = 3 s^2_{13}$, one can fit the values of $a^2$, $T^{}_1$, $T^{}_2$ and $T^{}_3$ as functions of the lightest neutrino mass by inputting the measured values of $s^2_{13}$, $\Delta m^2_{21}$ and $\Delta m^2_{31}$.

\begin{figure*}[t]
\centering
\includegraphics[width=6.5in]{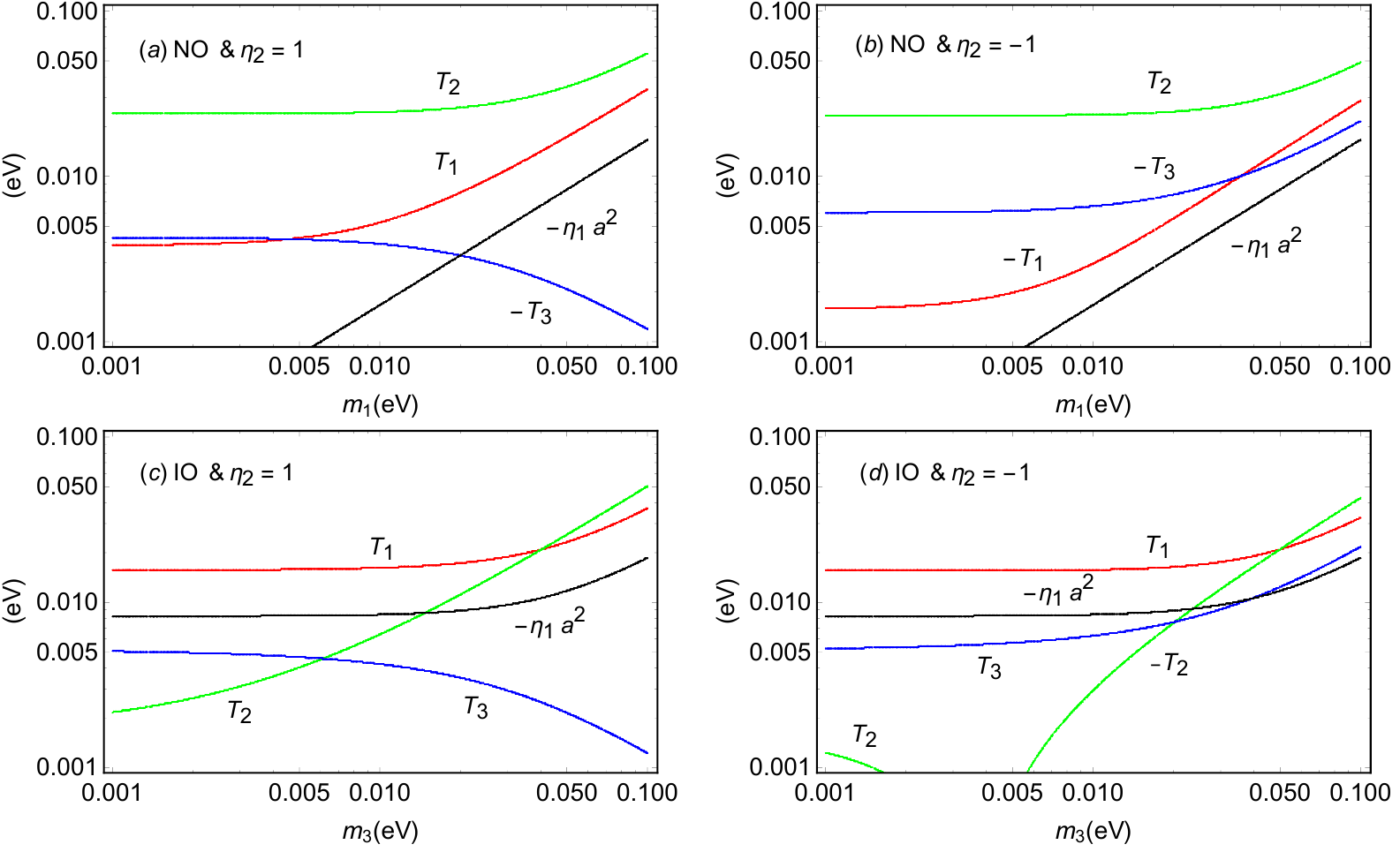}
\caption{ In the TM1 $\mu$-$\tau$ reflection symmetry scenario, the fitted values of $a^2$, $T^{}_1$, $T^{}_2$ and $T^{}_3$ as functions of the lightest neutrino mass in the NO and IO cases with $\eta^{}_2 = \pm 1$. }
\label{fig1}
\end{figure*}

Let us first consider the NO case. For $\eta^{}_2=1$, we have $T^{}_1, T^{}_2 >0$ and $T^2_3 - T^{}_1 T^{}_2 = -(be-cd)^2 <0$. As a direct result, both $m^{\prime}_2$ and $m^{\prime}_3$ in Eq.~(\ref{2.1.10}) are negative. And, in order to realize $|m^{\prime}_2| < |m^{\prime}_3|$, they should respectively take the ``$+$" and ``$-$" signs out of ``$\pm$". Taking the best-fit values of $s^2_{13}$, $\Delta m^2_{21}$ and $\Delta m^2_{31}$ as typical inputs, we plot the fitted values of $-\eta^{}_1 a^2$, $T^{}_1$, $T^{}_2$ and $-T^{}_3$ as functions of $m^{}_1$ in Fig.~\ref{fig1}(a). In obtaining these results we have taken $\delta = -\pi/2$, while the results for $\delta = \pi/2$ can be obtained by making the replacement $T^{}_3 \to - T^{}_3$. This is because the sign of $\delta$ coincides with that of $\sin \theta$ which is directly controlled by that of $T^{}_3$ [see Eqs.~(\ref{2.1.9}, \ref{2.1.12})]. The same goes for the following several cases to be discussed. One can convert $m^{\prime}_2$ and $m^{\prime}_3$ (and also $m^{\prime}_1$, in the case of $\eta^{}_1=1$) to positive by adding a diagonal phase matrix ${\rm diag}(\sqrt{-\eta^{}_1}, {\rm i}, {\rm i})$ to the right-hand side of $U^{}_0$ in Eq.~(\ref{2.1.8}). In this way the final lepton flavor mixing matrix turns out to be $U= U^{}_0 {\rm diag}(\sqrt{-\eta^{}_1}, {\rm i}, {\rm i})$. By utilizing the freedom of redefining the unphysical phases, such a $U$ can be converted into the following standard-parametrization form
\begin{eqnarray}
U = \displaystyle \frac{1}{\sqrt 6}
 \left( \begin{matrix}
2 & \sqrt{2} \cos \theta & - \sqrt{2} {\rm i} \sin \theta \cr
\displaystyle \frac{  - \sqrt{3} \cos \theta - {\rm i}  \sqrt{2}  \sin \theta }{\sqrt{3-\sin^2 \theta}}
& \displaystyle \frac{ \sqrt{6} - {\rm i}  \cos \theta \sin \theta  }{\sqrt{3-\sin^2 \theta}}  & \sqrt{3-\sin^2 \theta} \cr
\displaystyle \frac{  \sqrt{3} \cos \theta - {\rm i}  \sqrt{2}  \sin \theta  }{\sqrt{3-\sin^2 \theta}}
& \displaystyle \frac{ - \sqrt{6} - {\rm i}  \cos \theta \sin \theta   }{\sqrt{3-\sin^2 \theta}} &  \sqrt{3-\sin^2 \theta}
\end{matrix} \right) \left( \begin{matrix}
\sqrt{-\eta^{}_1} &  & \cr
& {\rm i}  & \cr
&  & 1
\end{matrix} \right) \;.
\label{2.1.11}
\end{eqnarray}
This not only explicitly reproduces the predictions of the TM1 mixing and $\mu$-$\tau$ reflection symmetry for the lepton flavor mixing parameters given in Eqs.~(\ref{4}, \ref{5}), but further tells us that \begin{eqnarray}
\delta = {\rm sign}(\sin \theta) \frac{\pi}{2} \;, \hspace{1cm} e^{{\rm i}\rho} = \sqrt{-\eta^{}_1}  \;, \hspace{1cm} \sigma = \frac{\pi}{2} \;.
\label{2.1.12}
\end{eqnarray}

Before proceeding, we would like to point out that since the measurable neutrino parameters only depend on three combinations (i.e., $T^{}_1, T^{}_2$ and $T^{}_3$) of the four model parameters $b$, $c$, $d$ and $e$, one is left with one degree of freedom in reconstructing the latter from the former: $T^{}_1, T^{}_2$ and $T^{}_3$ keep invariant under the following transformation
\begin{eqnarray}
\left( \begin{array}{c} b^\prime \cr d^\prime \end{array} \right) = \left( \begin{array}{cc} \cos \vartheta & -\sin \vartheta \cr \sin \vartheta & \cos \vartheta \end{array} \right) \left( \begin{array}{c} b \cr d \end{array} \right) \;, \hspace{1cm}
\left( \begin{array}{c} c^\prime \cr e^\prime \end{array} \right) = \left( \begin{array}{cc} \cos \vartheta & -\sin \vartheta \cr \sin \vartheta & \cos \vartheta \end{array} \right) \left( \begin{array}{c} c \cr e \end{array} \right) \;,
\label{2.1.13}
\end{eqnarray}
with $\vartheta$ being a real parameter.
With the help of such a degree of freedom, one can further reduce the model parameters by transforming one of $b$ and $d$ (or $c$ and $e$) to vanishing. To facilitate the following study on leptogenesis, we further identify this degree of freedom as one parameter of the Casas-Ibarra parametrization for $M^{}_{\rm D}$ \cite{CI}:
\begin{eqnarray}
M^{}_{\rm D} = {\rm i} U \sqrt{D^{}_\nu} R \sqrt{D^{}_{\rm R}}  \;,
\label{2.1.14}
\end{eqnarray}
with $R^T R = I$. In the case under consideration, by taking account of the results in Eqs.~(\ref{4}, \ref{5}, \ref{2.1.12}) and properly redefining the unphysical phases, one can transform ${\rm i} U$ to a form as
\begin{eqnarray}
{\rm i} U = \left( \begin{array}{ccc}
\displaystyle \frac{2}{\sqrt 6} &  s^{}_{12} c^{}_{13} & \pm s^{}_{13} \cr
\displaystyle \frac{1}{\sqrt 6}  & -  s^{}_{12} c^{}_{13} \mp {\rm i} \sqrt{\displaystyle \frac{3}{2}} s^{}_{13}  & \mp s^{}_{13} + {\rm i} \sqrt{\displaystyle \frac{3}{2}}  s^{}_{12} c^{}_{13} \cr
\displaystyle \frac{1}{\sqrt 6} & - s^{}_{12} c^{}_{13} \pm {\rm i} \sqrt{\displaystyle \frac{3}{2}}  s^{}_{13}  &  \mp s^{}_{13} -{\rm i} \sqrt{\displaystyle \frac{3}{2}}  s^{}_{12} c^{}_{13} \end{array} \right) \left( \begin{matrix}
\sqrt{\eta^{}_1} &  & \cr
& 1  & \cr
&  & 1
\end{matrix} \right) \;,
\label{2.1.15}
\end{eqnarray}
with $s^{}_{12} c^{}_{13} = \sqrt{(1-3 s^2_{13})/3}$. Here (and in the following) the upper and lower signs of $\pm$ and $\mp$ correspond to $\delta = -\pi/2$ and $\pi/2$, respectively.
Subsequently, the requirement of $M^{}_{\rm D}$ having a pattern described by Eq.~(\ref{2.1.5}) restricts $R$ to a form as
\begin{eqnarray}
R= \left( \begin{matrix}
1 & 0 & 0 \cr
0 & \cos \vartheta  & \sin \vartheta \cr
0 & - \sin \vartheta & \cos \vartheta
\end{matrix} \right) \;.
\label{2.1.16}
\end{eqnarray}

For $\eta^{}_2=-1$, because of $T^2_3 - T^{}_1 T^{}_2 = (ad-bc)^2 >0$, $m^{\prime}_2$ and $m^{\prime}_3$ in Eq.~(\ref{2.1.10}) now have opposite signs. In order to realize $|m^{\prime}_2| < |m^{\prime}_3|$, they should respectively take the ``$+$"  and ``$-$" (or ``$-$" and ``$+$") signs out of ``$\pm$" for $3T^{}_1 + 2 T^{}_2 >0$ (or $<0$).
In Fig.~\ref{fig1}(b), also for $\delta = -\pi/2$\footnote{ As in the previous case, the results for $\delta = \pi/2$ can be obtained by making the replacement $T^{}_3 \to - T^{}_3$.}, we plot the fitted values of $-\eta^{}_1 a^2$, $-T^{}_1$, $T^{}_2$ and $-T^{}_3$ as functions of $m^{}_1$ for $3T^{}_1 + 2 T^{}_2 >0$, while the results for $3T^{}_1 + 2 T^{}_2 <0$ just differ by a sign for $T^{}_1$, $T^{}_2$ and $T^{}_3$.
In the former (or latter) case, one has $m^{\prime}_2 >0 $ and $m^{\prime}_3 <0$ (or $m^{\prime}_2 <0 $ and $m^{\prime}_3 >0$), for which the final lepton flavor mixing matrix turns out to be $U= U^{}_0 {\rm diag}(\sqrt{-\eta^{}_1}, 1, {\rm i})$ (or $U^{}_0 {\rm diag}(\sqrt{-\eta^{}_1}, {\rm i}, 1)$), giving $e^{{\rm i}\rho}=\sqrt{-\eta^{}_1}$ (or $\sqrt{\eta^{}_1}$) and $\sigma =0$.

In the case under consideration, we are also left with one degree of freedom in reconstructing the model parameters $b$, $c$, $d$ and $e$ from the measurable neutrino parameters: $T^{}_1, T^{}_2$ and $T^{}_3$ keep invariant under the following transformation
\begin{eqnarray}
\left( \begin{array}{c} {\rm i} b^\prime \cr d^\prime \end{array} \right) = \left( \begin{array}{cc} \cosh \vartheta & -{\rm i} \sinh \vartheta \cr {\rm i} \sinh \vartheta & \cosh \vartheta \end{array} \right) \left( \begin{array}{c} {\rm i} b \cr d \end{array} \right) \;, \hspace{1cm}
\left( \begin{array}{c} {\rm i} c^\prime \cr e^\prime \end{array} \right) = \left( \begin{array}{cc} \cosh \vartheta & - {\rm i} \sinh \vartheta \cr {\rm i} \sinh \vartheta & \cosh \vartheta \end{array} \right) \left( \begin{array}{c} {\rm i} c \cr e \end{array} \right) \;.
\label{2.1.17}
\end{eqnarray}
This degree of freedom can also be identified as one parameter of the Casas-Ibarra parametrization for $M^{}_{\rm D}$: now ${\rm i}U$ has a form that only differs from that in Eq.~(\ref{2.1.15}) by a factor $-{\rm i}$  for the second column. Then, the requirement of $M^{}_{\rm D}$ having a pattern described by Eq.~(\ref{2.1.5}) restricts $R$ to a form as
\begin{eqnarray}
R= \left( \begin{matrix}
1 & 0 & 0 \cr
0 & \cosh \vartheta  & {\rm i} \sinh \vartheta \cr
0 & - {\rm i} \sinh \vartheta & \cosh \vartheta
\end{matrix} \right) \;.
\label{2.1.18}
\end{eqnarray}

The results for the IO case can be obtained in a similar way. For $\eta^{}_2=1$, in order to realize $|m^{\prime}_2| > |m^{\prime}_3|$, $m^{\prime}_2$ and $m^{\prime}_3$ in Eq.~(\ref{2.1.10}) should respectively take the ``$-$" and ``$+$" signs out of ``$\pm$". In Fig.~\ref{fig1}(c), also for $\delta = -\pi/2$, we plot the fitted values of $-\eta^{}_1 a^2$, $T^{}_1$, $T^{}_2$ and $T^{}_3$ as functions of $m^{}_3$.
In this case, the predictions for the lepton flavor mixing parameters are same as in the NO case.
And the $R$ matrix of the Casas-Ibarra parametrization for $M^{}_{\rm D}$ has a same form as in Eq.~(\ref{2.1.16}).

For $\eta^{}_2=-1$, in order to realize $|m^{\prime}_2| > |m^{\prime}_3|$, $m^{\prime}_2$ and $m^{\prime}_3$ in Eq.~(\ref{2.1.10}) should respectively take the ``$-$" and ``$+$" (or ``$+$" and ``$-$") signs out of ``$\pm$" for $3T^{}_1 + 2 T^{}_2 >0$ (or $<0$).
In Fig.~\ref{fig1}(d), also for $\delta = -\pi/2$, we plot the fitted values of $-\eta^{}_1 a^2$, $T^{}_1$, $|T^{}_2|$ and $T^{}_3$ as functions of $m^{}_3$ for $3T^{}_1 + 2 T^{}_2 >0$, while the results for $3T^{}_1 + 2 T^{}_2 <0$ just differ by a sign for $T^{}_1$, $T^{}_2$ and $T^{}_3$.
In the former (or latter) case, one has $m^{\prime}_2 <0 $ and $m^{\prime}_3 >0$ (or $m^{\prime}_2 >0 $ and $m^{\prime}_3 <0$), for which the final lepton flavor mixing matrix turns out to be $U= U^{}_0 {\rm diag}(\sqrt{-\eta^{}_1}, {\rm i}, 1)$ (or $U^{}_0 {\rm diag}(\sqrt{-\eta^{}_1}, 1, {\rm i})$), giving $e^{{\rm i}\rho}=\sqrt{\eta^{}_1}$ (or $\sqrt{-\eta^{}_1}$) and $\sigma=0$.
For both cases, the $R$ matrix of the Casas-Ibarra parametrization for $M^{}_{\rm D}$ has a same form as in Eq.~(\ref{2.1.18}).

Finally, for clarity, in Table~\ref{tab2} we summarize the predictions of different $(\eta^{}_1, \eta^{}_2)$ combinations for $\rho$ and $\sigma$ in the NO and IO cases.

\begin{table}\centering
  \begin{footnotesize}
    \begin{tabular}{c|cc|cc}
     \hline\hline
      & \multicolumn{2}{c|}{Normal Ordering}
      & \multicolumn{2}{c}{Inverted Ordering }
      \\
      \cline{2-5}
      & $3 T^{}_1 + 2 T^{}_2 >0$ & $3 T^{}_1 + 2 T^{}_2 <0$
      & $3 T^{}_1 + 2 T^{}_2 >0$ & $3 T^{}_1 + 2 T^{}_2 <0$
      \\
      \cline{1-5}
      \rule{0pt}{4mm}\ignorespaces
       (1, 1)  & ($\pi/2$, $\pi/2$) & $-$ & ($\pi/2$, $\pi/2$) &  $-$
      \\[1mm]
       ($-1$, 1) & ($0$, $\pi/2$) &   $-$ & ($0$, $\pi/2$) &  $-$
      \\[1mm]
      (1, $-1$)  & ($\pi/2$, $0$) & ($0$, $0$)  & ($0$, $0$)  &  ($\pi/2$, $0$)
      \\[1mm]
     ($-1$, $-1$) & ($0$, $0$)  &  ($\pi/2$, $0$) & ($\pi/2$, $0$) &   ($0$, $0$)
      \\[2mm]
      \hline\hline
    \end{tabular}
  \end{footnotesize}
  \caption{In the TM1 $\mu$-$\tau$ reflection symmetry scenario, for $3 T^{}_1 + 2 T^{}_2 >0$ and $<0$, the predicted values of $(\rho, \sigma)$ in the NO and IO cases with $(\eta^{}_1, \eta^{}_2) = (1, 1), (-1, 1), (1, -1)$ and $(-1, -1)$. }
\label{tab2}
\end{table}

\subsection{TM2 $\mu$-$\tau$ reflection symmetry in the seesaw model}

In this subsection, we study the combination of the TM2 mixing and $\mu$-$\tau$ reflection symmetry. In the seesaw model, the TM2 mixing can be naturally realized by taking one column of $M^{}_{\rm D}$ to be proportional to the second column of $U^{}_{\rm TBM}$ and the other two columns to be orthogonal to it \cite{TMss}:
\begin{eqnarray}
M^{}_{\rm D}= \left( \begin{array}{ccc}
2 b \sqrt{M^{}_1} & a \sqrt{M^{}_2} &   2 d \sqrt{M^{}_3} \cr
(b-c) \sqrt{M^{}_1} & -a \sqrt{M^{}_2} &   (d-e) \sqrt{M^{}_3} \cr
(b+c)  \sqrt{M^{}_1}  & -a \sqrt{M^{}_2}  &   (d+e)  \sqrt{M^{}_3} \cr
\end{array} \right) \;,
\label{2.2.1}
\end{eqnarray}
where $a$, $b$, $c$, $d$ and $e$ are generally complex. One can directly verify that the resulting $M^{}_\nu$ from such an $M^{}_{\rm D}$ does obey the TM2 symmetry:
\begin{eqnarray}
M^{}_\nu = R^{}_{\rm TM2} M^{}_\nu R^{}_{\rm TM2}
\hspace{0.5cm} {\rm with} \hspace{0.5cm}
R^{}_{\rm TM2}= - \frac{1}{3} \left( \begin{array}{ccc}
1 & 2 & 2 \cr
2 & 1 & -2 \cr
2 & -2 & 1 \cr
\end{array} \right)  \;.
\label{2.2.2}
\end{eqnarray}
Then, it is easy to see that, in order for $M^{}_{\rm D}$ to be consistent with the patterns described by Eqs.~(\ref{2.1.3}, \ref{2.2.1}) simultaneously, it should take a form as
\begin{eqnarray}
M^{}_{\rm D}= \left( \begin{array}{ccc}
 2 b \sqrt{M^{}_1} &  a \sqrt{M^{}_2} & 2 d \sqrt{M^{}_3} \cr
(b-{\rm i} c) \sqrt{M^{}_1} & -a \sqrt{M^{}_2} &    (d-{\rm i} e) \sqrt{M^{}_3} \cr
(b+{\rm i}c)  \sqrt{M^{}_1}  &  - a \sqrt{M^{}_2}  &  (d+{\rm i} e)  \sqrt{M^{}_3} \cr
\end{array} \right) P^{}_N \;,
\label{2.2.3}
\end{eqnarray}
with $a, b, c, d$ and $e$ being real.

The resulting $M^{}_\nu$ from $M^{}_{\rm D}$ in Eq.~(\ref{2.2.3}) is given by
\begin{eqnarray}
&& M^{}_\nu = - \left( \begin{array}{ccc}
\eta^{}_2 a^2+ 4 T^{}_1 & - \eta^{}_2 a^2 +2 T^{}_1 - 2 {\rm i}T^{}_3 & - \eta^{}_2 a^2 +2 T^{}_1 + 2 {\rm i}T^{}_3 \cr
- \eta^{}_2 a^2 +2 T^{}_1 - 2{\rm i} T^{}_3 & \eta^{}_2 a^2 + T^{}_1 - T^{}_2 - 2 {\rm i} T^{}_3 & \eta^{}_2 a^2 + T^{}_1 + T^{}_2  \cr
- \eta^{}_2 a^2 +2 T^{}_1 + 2 {\rm i}T^{}_3  & \eta^{}_2 a^2 + T^{}_1 + T^{}_2  & \eta^{}_2 a^2 + T^{}_1 - T^{}_2 + 2 {\rm i} T^{}_3 \cr
\end{array} \right) \;,
\label{2.2.4}
\end{eqnarray}
where the definitions of $T^{}_1$, $T^{}_2$ and $T^{}_3$ are same as in Eq.~(\ref{2.1.7}) except for the replacement $\eta^{}_2 \to \eta^{}_1$.
One can directly verify that such an $M^{}_\nu$ does obey the TM2 and $\mu$-$\tau$ reflection symmetries simultaneously. This $M^{}_\nu$ can be diagonalized by the following unitary matrix
\begin{eqnarray}
U^{}_0= \displaystyle \frac{1}{\sqrt 6} \left( \begin{array}{ccc} \vspace{0.15cm}
2 & \sqrt{2} & 0 \cr \vspace{0.15cm}
1 & - \sqrt{2}  & -\sqrt{3}  \cr
1 & - \sqrt{2}  & \sqrt{3} \cr
\end{array} \right)  \left( \begin{array}{ccc}
1 & 0 & 0 \cr
0 & 1  & 0  \cr
0 & 0  & {\rm i} \cr
\end{array} \right) \left( \begin{array}{ccc}
\cos \theta & 0 & \sin \theta \cr
0 &  1 & 0  \cr
 -\sin \theta  & 0 & \cos \theta \cr
\end{array} \right) \; ,
\label{2.2.5}
\end{eqnarray}
with
\begin{eqnarray}
\tan 2 \theta = \frac{ 2\sqrt{3} T^{}_3 }{T^{}_2 - 3 T^{}_1} \;,
\label{2.2.6}
\end{eqnarray}
giving the neutrino masses as
\begin{eqnarray}
m^{\prime}_2= - 3 \eta^{}_2 a^2 \;, \hspace{1cm} m^{\prime}_{1, 3} =  -(3T^{}_1 + T^{}_2) \pm \sqrt{ ( 3T^{}_1 + T^{}_2)^2 + 12 (T^{2}_3 -T^{}_1 T^{}_2 )  }   \;.
\label{2.2.7}
\end{eqnarray}
With the help of Eqs.~(\ref{2.2.6}, \ref{2.2.7}) and $\sin^2 \theta = 3 s^2_{13}/2$, one can fit the values of $a^2$, $T^{}_1$, $T^{}_2$ and $T^{}_3$ as functions of the lightest neutrino mass by inputting the measured values of $s^2_{13}$, $\Delta m^2_{21}$ and $\Delta m^2_{31}$.

\begin{figure*}[t]
\centering
\includegraphics[width=6.5in]{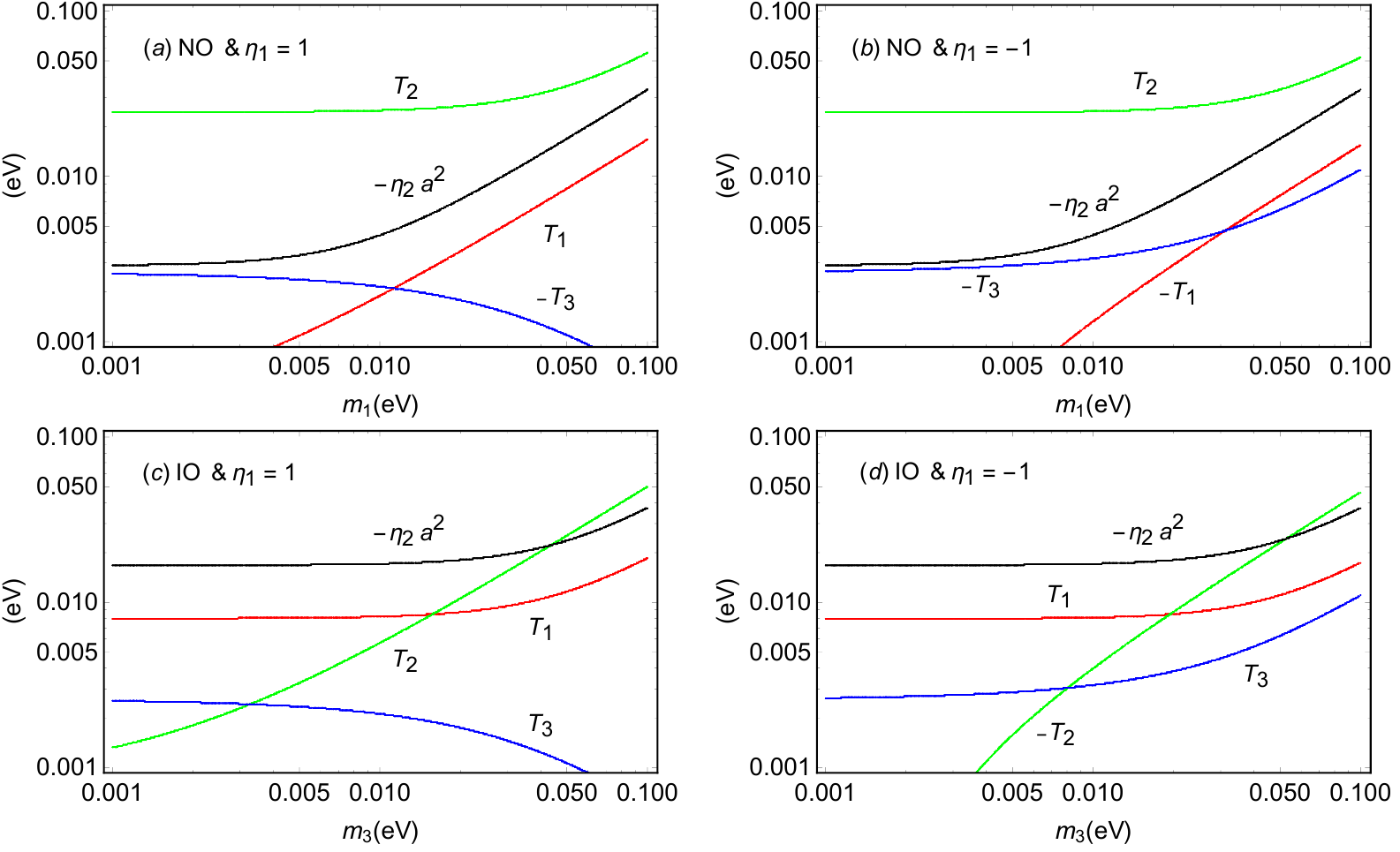}
\caption{ In the TM2 $\mu$-$\tau$ reflection symmetry scenario, the fitted values of $a^2$, $T^{}_1$, $T^{}_2$ and $T^{}_3$ as functions of the lightest neutrino mass in the NO and IO cases with $\eta^{}_1 = \pm 1$. }
\label{fig2}
\end{figure*}

Let us first consider the NO case. For $\eta^{}_1=1$, we have $T^{}_1, T^{}_2 >0$ and $T^2_3 - T^{}_1 T^{}_2 = -(be-cd)^2 <0$. As a direct result, both $m^{\prime}_1$ and $m^{\prime}_3$ in Eq.~(\ref{2.2.7}) are negative. In order to realize $|m^{\prime}_1| < |m^{\prime}_3|$, they should respectively take the ``$+$" and ``$-$" signs out of ``$\pm$". In Fig.~\ref{fig2}(a), we plot the fitted values of $-\eta^{}_2 a^2$, $T^{}_1$, $T^{}_2$ and $-T^{}_3$ as functions of $m^{}_1$. In obtaining these results we have also taken $\delta = -\pi/2$, while the results for $\delta = \pi/2$ can also be obtained by making the replacement $T^{}_3 \to - T^{}_3$. This is also because the sign of $\delta$ coincides with that of $\sin \theta$ which is directly controlled by that of $T^{}_3$ [see Eqs.~(\ref{2.2.6}, \ref{2.2.9})]. The same goes for the following several cases to be discussed.
One can convert $m^{\prime}_1$ and $m^{\prime}_3$ (and also $m^{\prime}_2$, in the case of $\eta^{}_2=1$) to positive by adding a diagonal phase matrix ${\rm diag}({\rm i}, \sqrt{-\eta^{}_2}, {\rm i})$ to the right-hand side of $U^{}_0$ in Eq.~(\ref{2.2.5}). In this way the final lepton flavor mixing matrix turns out to be $U= U^{}_0 {\rm diag}({\rm i}, \sqrt{-\eta^{}_2}, {\rm i})$. By utilizing the freedom of redefining the unphysical phases, such a $U$ can be converted into the following standard-parametrization form
\begin{eqnarray}
U  = \displaystyle \frac{1}{\sqrt 6}
\left( \begin{matrix}
2 \cos \theta & \sqrt{2}  & - 2 {\rm i} \sin \theta \cr
\displaystyle \frac{ - \sqrt{3}  - 2 {\rm i} \cos \theta  \sin \theta  }{\sqrt{3-2 \sin^2 \theta}}
& \displaystyle \frac{ \sqrt{6} \cos \theta - \sqrt{2} {\rm i} \sin \theta   }{\sqrt{3-2 \sin^2 \theta}}  & \sqrt{3-2 \sin^2 \theta} \cr
\displaystyle \frac{  \sqrt{3}  - 2 {\rm i} \cos \theta  \sin \theta  }{\sqrt{3-2 \sin^2 \theta}}
& \displaystyle \frac{ - \sqrt{6} \cos \theta - \sqrt{2} {\rm i} \sin \theta   }{\sqrt{3-2 \sin^2 \theta}} &  \sqrt{3-2\sin^2 \theta}
\end{matrix} \right) \left( \begin{matrix}
{\rm i} &  & \cr
& \sqrt{-\eta^{}_2}  & \cr
&  & 1
\end{matrix} \right) \;.
\label{2.2.8}
\end{eqnarray}
This not only explicitly reproduces the predictions of the TM2 mixing and $\mu$-$\tau$ reflection symmetry for the lepton flavor mixing parameters given in Eqs.~(\ref{4}, \ref{5}), but further tells us that
\begin{eqnarray}
\delta = {\rm sign}(\sin \theta) \frac{\pi}{2} \;, \hspace{1cm}  \rho = \frac{\pi}{2}  \;, \hspace{1cm} e^{{\rm i}\sigma} = \sqrt{-\eta^{}_2}  \;.
\label{2.2.9}
\end{eqnarray}
Then, we consider the Casas-Ibarra parametrization for $M^{}_{\rm D}$ in the case under consideration: by taking account of the results in Eqs.~(\ref{4}, \ref{5}, \ref{2.2.9}) and properly redefining the unphysical phases, one can transform ${\rm i} U$ to a form as
\begin{eqnarray}
{\rm i} U =  \left( \begin{array}{ccc}
c^{}_{12} c^{}_{13} & \displaystyle \frac{1}{\sqrt 3} & \pm s^{}_{13} \cr
\displaystyle \frac{1}{2} c^{}_{12} c^{}_{13} \mp  {\rm i} \displaystyle \frac{\sqrt 3}{ 2} s^{}_{13} & - \displaystyle  \frac{1}{\sqrt 3}  & \pm \displaystyle \frac{1}{2} s^{}_{13} + {\rm i} \displaystyle \frac{\sqrt 3}{ 2}  c^{}_{12} c^{}_{13} \cr
\displaystyle \frac{1}{2} c^{}_{12} c^{}_{13} \pm  {\rm i} \displaystyle \frac{\sqrt 3}{ 2} s^{}_{13} & - \displaystyle  \frac{1}{\sqrt 3}  & \pm \displaystyle \frac{1}{2} s^{}_{13} - {\rm i} \displaystyle \frac{\sqrt 3}{ 2}  c^{}_{12} c^{}_{13}
\end{array} \right) \left( \begin{matrix}
1 &  & \cr
& \sqrt{\eta^{}_2}  & \cr
&  & 1
\end{matrix} \right) \;,
\label{2.2.10}
\end{eqnarray}
with $c^{}_{12} c^{}_{13} = \sqrt{(2-3 s^2_{13})/3}$. Subsequently, the requirement of $M^{}_{\rm D}$ having a pattern described by Eq.~(\ref{2.2.3}) restricts $R$ to a form as
\begin{eqnarray}
R= \left( \begin{matrix}
\cos \vartheta & 0 & \sin \vartheta \cr
0 &  1 & 0 \cr
- \sin \vartheta & 0 & \cos \vartheta
\end{matrix} \right) \;.
\label{2.2.11}
\end{eqnarray}

For $\eta^{}_1=-1$, because of $T^2_3 - T^{}_1 T^{}_2 = (ad-bc)^2 >0$, $m^{\prime}_1$ and $m^{\prime}_3$ in Eq.~(\ref{2.2.7}) now have opposite signs. In order to realize $|m^{\prime}_1| < |m^{\prime}_3|$, they should respectively take the ``$+$" and ``$-$" (or ``$-$" and ``$+$") signs out of ``$\pm$" for $3T^{}_1 + T^{}_2 >0$ (or $<0$).
In Fig.~\ref{fig2}(b), also for $\delta = -\pi/2$, we plot the fitted values of $-\eta^{}_2 a^2$, $-T^{}_1$, $T^{}_2$ and $-T^{}_3$ as functions of $m^{}_1$ for $3T^{}_1 + T^{}_2 >0$, while the results for $3T^{}_1 + T^{}_2 <0$ just differ by a sign for $T^{}_1$, $T^{}_2$ and $T^{}_3$. In the former (or latter) case, one has $m^{\prime}_1 >0 $ and $m^{\prime}_3 <0$ (or $m^{\prime}_1 <0 $ and $m^{\prime}_3 >0$), for which the final lepton flavor mixing matrix turns out to be $U= U^{}_0 {\rm diag}(1, \sqrt{-\eta^{}_2},  {\rm i})$ (or $U^{}_0 {\rm diag}({\rm i}, \sqrt{-\eta^{}_2}, 1)$), giving $\rho =0$ and $e^{{\rm i}\sigma}=\sqrt{-\eta^{}_2}$ (or $\sqrt{\eta^{}_2}$).
In the case under consideration, the Casas-Ibarra parametrization for $M^{}_{\rm D}$ goes as follows: now ${\rm i}U$ has a form that only differs from that in Eq.~(\ref{2.2.10}) by a factor $-{\rm i}$ for the first column. Then, the requirement of $M^{}_{\rm D}$ having a pattern described by Eq.~(\ref{2.2.3}) restricts $R$ to a form as
\begin{eqnarray}
R= \left( \begin{matrix}
\cosh \vartheta & 0 & {\rm i} \sinh \vartheta \cr
0 &  1 &  0 \cr
- {\rm i} \sinh \vartheta & 0 & \cosh \vartheta
\end{matrix} \right) \;.
\label{2.2.12}
\end{eqnarray}

The results for the IO case can be obtained in a similar way. For $\eta^{}_1=1$, in order to realize $|m^{\prime}_1| > |m^{\prime}_3|$, $m^{\prime}_1$ and $m^{\prime}_3$ in Eq.~(\ref{2.2.7}) should respectively take the ``$-$" and ``$+$" signs out of ``$\pm$". In Fig.~\ref{fig2}(c), also for $\delta = -\pi/2$, we plot the fitted values of $-\eta^{}_2 a^2$, $T^{}_1$, $T^{}_2$ and $T^{}_3$ as functions of $m^{}_3$.
In the present case, the results for the lepton flavor mixing parameters are same as those in the NO case. And the $R$ matrix of the Casas-Ibarra parametrization for $M^{}_{\rm D}$ has a same form as in Eq.~(\ref{2.2.11}).

For $\eta^{}_1=-1$, in order to realize $|m^{\prime}_1| > |m^{\prime}_3|$, $m^{\prime}_1$ and $m^{\prime}_3$ in Eq.~(\ref{2.2.7}) should respectively take the ``$-$" and ``$+$" (or `$+$" and ``$-$") signs out of ``$\pm$" for $3T^{}_1 + 2 T^{}_2 >0$ (or $<0$).
In Fig.~\ref{fig2}(d), also for $\delta = -\pi/2$, we plot the fitted values of $-\eta^{}_2 a^2$, $T^{}_1$, $T^{}_2$ and $T^{}_3$ as functions of $m^{}_3$ for $3T^{}_1 + T^{}_2 >0$, while the results for $3T^{}_1 + T^{}_2 <0$ just differ by a sign for $T^{}_1$, $T^{}_2$ and $T^{}_3$.
In the former (or latter) case, one has $m^{\prime}_1 <0 $ and $m^{\prime}_3 >0$ (or $m^{\prime}_1 >0 $ and $m^{\prime}_3 <0$), for which the final lepton flavor mixing matrix turns out to be $U= U^{}_0 {\rm diag}( {\rm i}, \sqrt{-\eta^{}_2}, 1)$ (or $U^{}_0 {\rm diag}(1, \sqrt{-\eta^{}_2}, {\rm i})$), giving $\rho =0$ and $e^{{\rm i}\sigma}=\sqrt{\eta^{}_2}$ (or $\sqrt{-\eta^{}_2}$).
For both cases, the $R$ matrix of the Casas-Ibarra parametrization for $M^{}_{\rm D}$ has a same form as in Eq.~(\ref{2.2.12}).

Finally, for clarity, in Table~\ref{tab3} we summarize the predictions of different $(\eta^{}_1, \eta^{}_2)$ combinations for $\rho$ and $\sigma$ in the NO and IO cases.

\begin{table}\centering
  \begin{footnotesize}
    \begin{tabular}{c|cc|cc}
     \hline\hline
      & \multicolumn{2}{c|}{Normal Ordering}
      & \multicolumn{2}{c}{Inverted Ordering }
      \\
      \cline{2-5}
      & $3 T^{}_1 + T^{}_2 >0$ & $3 T^{}_1 + T^{}_2 <0$
      & $3 T^{}_1 + T^{}_2 >0$ & $3 T^{}_1 + T^{}_2 <0$
      \\
      \cline{1-5}
      \rule{0pt}{4mm}\ignorespaces
      (1, 1)  & ($\pi/2$, $\pi/2$) & $-$ & ($\pi/2$, $\pi/2$) &  $-$
      \\[1mm]
      (1, $-1$)  & ($\pi/2$, $0$) & $-$  & ($\pi/2$, $0$)  &  $-$
      \\[1mm]
      ($-1$, 1) & ($0$, $\pi/2$) & ($0$, $0$)  & ($0$, $0$) &  ($0$, $\pi/2$)
      \\[1mm]
     ($-1$, $-1$) & ($0$, $0$)  &  ($0$, $\pi/2$) &  ($0$, $\pi/2$)  &  ($0$, $0$)
      \\[2mm]
      \hline\hline
    \end{tabular}
  \end{footnotesize}
  \caption{In the TM2 $\mu$-$\tau$ reflection symmetry scenario, for $3 T^{}_1 + T^{}_2 >0$ and $<0$, the predicted values of $(\rho, \sigma)$ in the NO and IO cases with $(\eta^{}_1, \eta^{}_2) = (1, 1), (-1, 1), (1, -1)$ and $(-1, -1)$. }
\label{tab3}
\end{table}

\subsection{A possible approach to get the desired mass matrices}

In this subsection, we discuss a possible approach to get the textures of $M^{}_{\rm D}$ in Eqs.~(\ref{2.1.5}, \ref{2.2.3}) responsible for the trimaximal $\mu$-$\tau$ reflection symmetry. In the literature, the TBM mixing can be realized in a way as follows \cite{FS}: a particular flavor symmetry such as the $A^{}_4$ or $S^{}_4$ group which contains triplet representations is invoked. And three lepton doublets constitute a triplet representation under it while three right-handed neutrinos are simply singlet representations. In order to properly break the flavor symmetry and generate the desired neutrino masses, three so-called flavon fields $\phi^{}_{J}$ (for $J=1, 2, 3$) are introduced, each of which is a triplet representation [with three components $\phi^{}_{J} = (\phi^{}_{J1}, \phi^{}_{J2}, \phi^{}_{J3})^T$] under the flavor symmetry. Under such a setup, after the Higgs and flavon fields acquire non-vanishing vacuum expectation values (VEVs), the neutrino masses will arise from the following operators
\begin{eqnarray}
\sum^{}_{I, J} \frac{ y^{}_{IJ}}{\Lambda}[ \overline L^{}_e \phi^{}_{J1} + \overline L^{}_\mu \phi^{}_{J2} + \overline L^{}_\tau \phi^{}_{J3} ] H N^{}_I \;,
\label{2.3.1}
\end{eqnarray}
where $y^{}_{IJ}$ are coefficients and $\Lambda$ stands for the cutoff scale. When three flavon fields are respectively associated with three right-handed neutrinos (i.e., $y^{}_{IJ} = 0$ for $I \neq J$) and develop the following particular VEV alignments
\begin{eqnarray}
&& (\langle \phi^{}_{11} \rangle, \langle \phi^{}_{12} \rangle, \langle \phi^{}_{13} \rangle) \propto (2, 1, 1)^T \;, \nonumber \\
&& (\langle \phi^{}_{21} \rangle, \langle \phi^{}_{22} \rangle, \langle \phi^{}_{23} \rangle)  \propto (1, -1, -1)^T \;, \nonumber \\
&& (\langle \phi^{}_{31} \rangle, \langle \phi^{}_{32} \rangle, \langle \phi^{}_{33} \rangle)  \propto (0, -1, 1)^T \;,
\label{2.3.2}
\end{eqnarray}
$M^{}_{\rm D}$ will take a form as
\begin{eqnarray}
M^{}_{\rm D}= \left( \begin{array}{ccc} \vspace{0.15cm}
2 a & b  & 0 \cr \vspace{0.15cm}
a & -b & -c \cr
a & - b & c \cr
\end{array} \right)  \;,
\label{2.3.3}
\end{eqnarray}
and subsequently lead the neutrino mixing to be of the TBM pattern.

By slightly modifying the above approach to get the texture of $M^{}_{\rm D}$ responsible for the TBM mixing, the textures of $M^{}_{\rm D}$ in Eqs.~(\ref{2.1.5}, \ref{2.2.3}) responsible for the trimaximal $\mu$-$\tau$ reflection symmetry can be similarly realized: the CP symmetry will be employed in combination with the flavor symmetry, which will restrict all the coefficients to be real. The VEV alignments of three flavon fields are same as in Eq.~(\ref{2.3.2}). But the VEV of $\phi^{}_3$ is purely imaginary now: $(\langle \phi^{}_{31} \rangle, \langle \phi^{}_{32} \rangle, \langle \phi^{}_{33} \rangle)  \propto {\rm i} (0, -1, 1)^T$, which spontaneously breaks the CP symmetry so as to generate non-trivial CP phases. Then, the texture of $M^{}_{\rm D}$ in Eq.~(\ref{2.1.5}) [and similarly for that in Eq.~(\ref{2.2.3})] will arise if one associates $\phi^{}_1$ with $N^{}_1$ and both $\phi^{}_2$ and $\phi^{}_3$ with $N^{}_2$ and $N^{}_3$.

\subsection{Compatibility with texture zeros}

\begin{figure*}[t]
\centering
\includegraphics[width=6.5in]{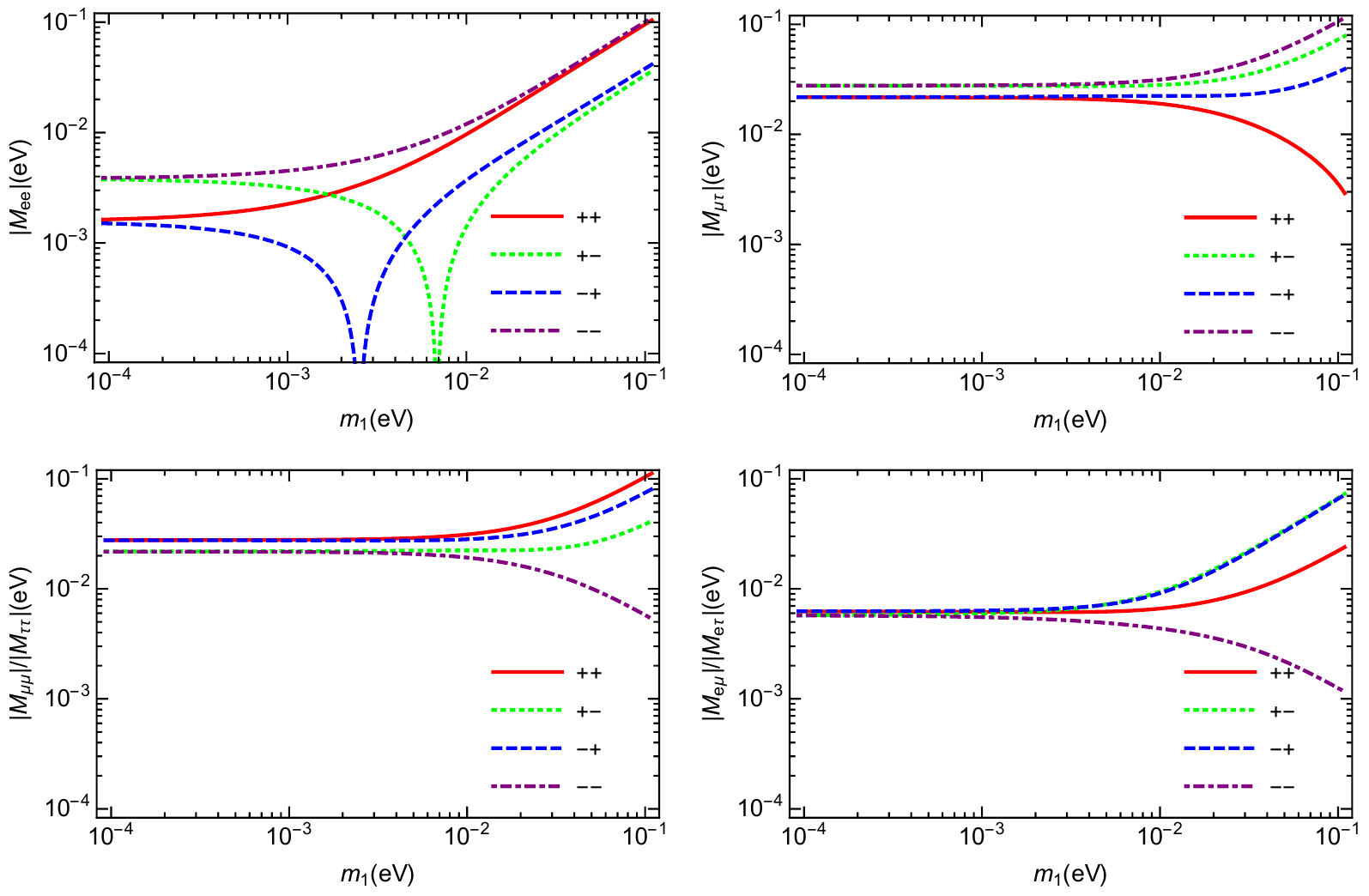}
\caption{ In the TM1 $\mu$-$\tau$ reflection symmetry scenario, $|M^{}_{\alpha \beta}|$ as functions of the lightest neutrino mass $m^{}_1$ in the NO case for $(\eta^{}_\rho, \eta^{}_\sigma) = (1, 1), (1, -1), (-1, 1)$ and $(-1, -1)$ (denoted as $++$, $-+$, $+-$ and $--$, respectively). }
\label{fig3}
\end{figure*}

\begin{figure*}[t]
\centering
\includegraphics[width=6.5in]{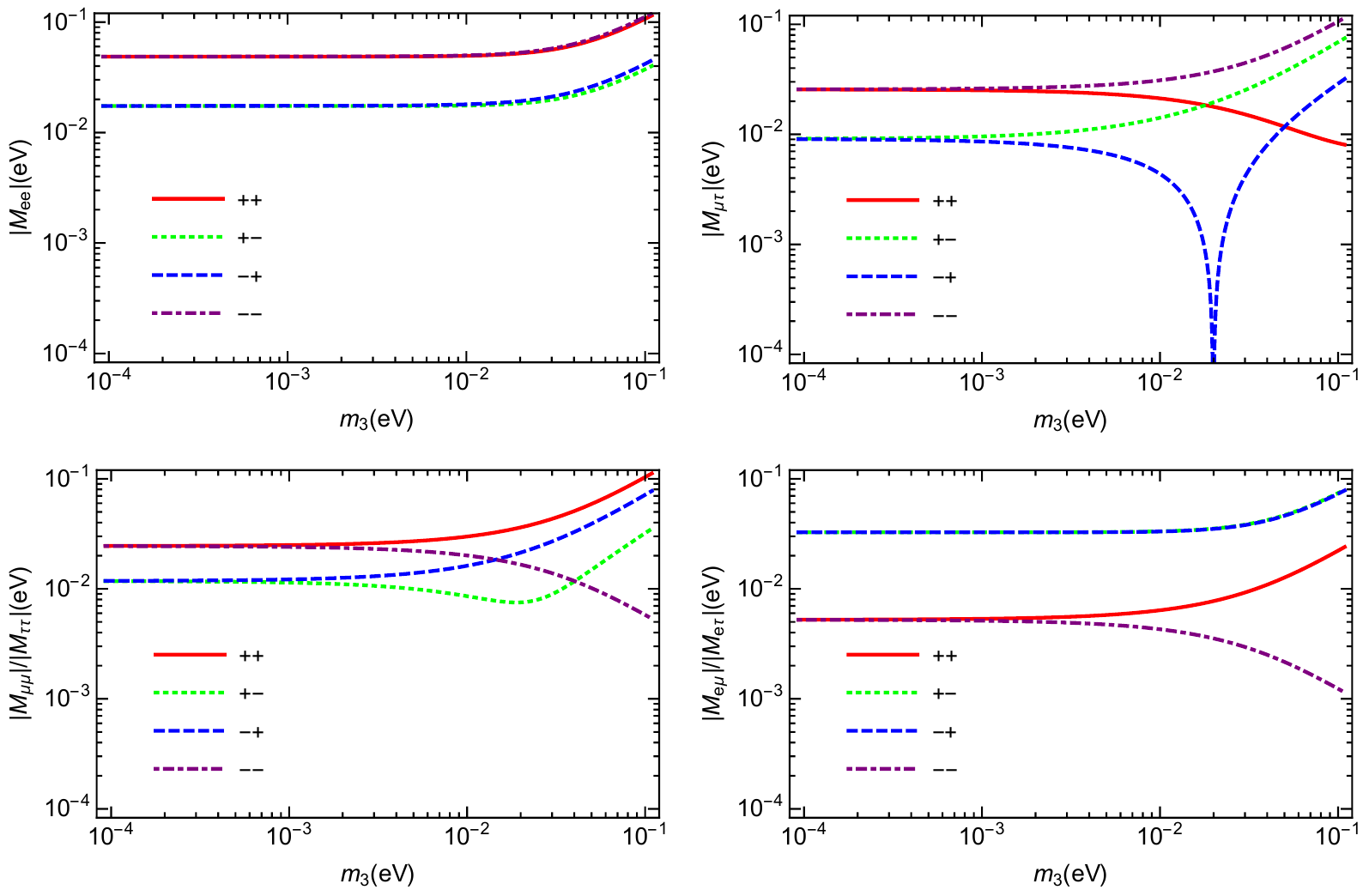}
\caption{ In the TM1 $\mu$-$\tau$ reflection symmetry scenario, $|M^{}_{\alpha \beta}|$ as functions of the lightest neutrino mass $m^{}_3$ in the IO case for $(\eta^{}_\rho, \eta^{}_\sigma) = (1, 1), (1, -1), (-1, 1)$ and $(-1, -1)$ (denoted as $++$, $-+$, $+-$ and $--$, respectively). }
\label{fig4}
\end{figure*}

In this subsection, we study the compatibility of the trimaximal $\mu$-$\tau$ reflection symmetry with texture zeros of $M^{}_\nu$ --- an interesting possibility from the phenomenological point of view.
For the trimaximal $\mu$-$\tau$ reflection symmetry scenario, taking account of the predictions for the lepton flavor mixing parameters, the magnitudes of the elements of $M^{}_\nu$ are directly obtained as
\begin{eqnarray}
| M^{}_{ee} |  & = & \left| m^{}_1 \eta^{}_\rho c^2_{12} c^2_{13}+
m^{}_2  \eta^{}_\sigma s^2_{12} c^2_{13} - m^{}_3 s^{2}_{13} \right| \; ,
\nonumber \\
| M^{}_{\mu\tau} | & = &  \frac{1}{2}  \left|  m^{}_1 \eta^{}_\rho \left(1
- c^{2}_{12} c^{2}_{13}  \right)
+ m_2^{} \eta^{}_\sigma \left(1- s^{2}_{12} c^{2}_{13} \right)
-  m^{}_3 c^2_{13}  \right| \; ,
\nonumber \\
| M^{}_{\mu\mu} | = | M^{}_{\tau\tau} | & = & \frac{1}{2}  \left| m^{}_1 \eta^{}_\rho \left(s^{}_{12}
\mp {\rm i} c^{}_{12} s^{}_{13} \right)^2
+ m^{}_2 \eta^{}_\sigma \left(c^{}_{12} \pm {\rm i} s^{}_{12}
s^{}_{13} \right)^2 + m^{}_3 c^2_{13} \right| \; ,
\nonumber \\
| M^{}_{e\mu} | = | M^{}_{e\tau} | & = & \frac{1}{\sqrt 2} c^{}_{13} \left| m^{}_1 \eta^{}_\rho c^{}_{12}  \left(s^{}_{12} \mp {\rm i} c^{}_{12} s^{}_{13} \right)
-  m^{}_2 \eta^{}_\sigma s^{}_{12} \left( c^{}_{12} \pm {\rm i}
s^{}_{12} s^{}_{13} \right)  \mp {\rm i} m^{}_3 s^{}_{13} \right| \; ,
\label{2.4.1}
\end{eqnarray}
with $\eta^{}_\rho =1$ or $-1$ for $\rho=0$ or $\pi/2$ (and similarly for $\eta^{}_\sigma$) and $c^2_{12} c^2_{13} =2/3$ (or $s^2_{12} c^2_{13} =1/3$) for the TM1 (or TM2) case. One can see that the results of $|M^{}_{\alpha \beta}|$ are same for $\delta  =\pi/2$ and $-\pi/2$.
In Figs.~\ref{fig3} and \ref{fig4}, we plot
$|M^{}_{\alpha \beta}|$ as functions of the lightest neutrino mass in the NO and IO cases
for $(\eta^{}_\rho, \eta^{}_\sigma) = (1, 1), (1, -1), (-1, 1)$ and $(-1, -1)$. Considering that the predictions of the TM1 case for the lepton flavor mixing parameters just differ from those of the TM2 case by a small difference for $\theta^{}_{12}$, here we have only shown the results for the former case. From these results we see that in the NO case $|M^{}_{ee}|$ may become vanishing at $m^{}_1 \simeq 2.5$ meV (or 6.8 meV) for $(\eta^{}_\rho, \eta^{}_\sigma) = (-1, 1)$ (or $(1, -1)$), while in the IO case $|M^{}_{\mu\tau}|$ may become vanishing at $m^{}_3 \simeq 0.02$ eV for $(\eta^{}_\rho, \eta^{}_\sigma) = (-1, 1)$.

\section{Implications for leptogenesis}

In this section, we study the implications of the trimaximal $\mu$-$\tau$ reflection symmetry for leptogenesis. As we know, the seesaw model can also provide an attractive explanation  for the baryon-antibaryon asymmetry of the Universe \cite{planck}
\begin{eqnarray}
Y^{}_{\rm B} \equiv \frac{n^{}_{\rm B}-n^{}_{\rm \bar B}}{s} = (8.67 \pm 0.15) \times 10^{-11}  \;,
\label{3.1}
\end{eqnarray}
where $n^{}_{\rm B}$ ($n^{}_{\rm \bar B}$) denotes the baryon (antibaryon) number density and $s$ the entropy density. This is just the leptogenesis mechanism \cite{leptogenesis,Lreview}: an asymmetry between matter and antimatter is firstly generated in the lepton sector (i.e., the lepton asymmetry $Y^{}_{\rm L} \equiv (n^{}_{\rm L}-n^{}_{\rm \bar L})/s$) and then converted to the baryon sector by $Y^{}_{\rm B} \simeq - c Y^{}_{\rm L}$ with $c= 28/79$. The decay processes of the right-handed neutrino $N^{}_I$ can produce a baryon asymmetry as
\begin{eqnarray}
Y^{}_{\rm B} = -c r \varepsilon^{}_I \kappa(\widetilde m^{}_I)  \;.
\label{3.2}
\end{eqnarray}
Here $r \simeq 3.9 \times 10^{-3}$ is the ratio of the number density of $N^{}_I$ to the entropy density. $\varepsilon^{}_I$ is the CP asymmetry for the decay processes of $N^{}_I$
\begin{eqnarray}
\varepsilon^{}_{I}
 =  \frac{1}{8\pi (M^\dagger_{\rm D}
M^{}_{\rm D})^{}_{II} v^2} \sum^{}_{J \neq I} {\rm Im}\left[ (M^\dagger_{\rm D} M^{}_{\rm D})^{2}_{IJ}\right] {\cal F} \left( \frac{M^2_J}{M^2_I} \right) \;,
\label{3.3}
\end{eqnarray}
which is a sum of the flavored CP asymmetries
\begin{eqnarray}
\varepsilon^{}_{I \alpha}
& = & \frac{1}{8\pi (M^\dagger_{\rm D}
M^{}_{\rm D})^{}_{II} v^2} \sum^{}_{J \neq I} \left\{ {\rm Im}\left[(M^*_{\rm D})^{}_{\alpha I} (M^{}_{\rm D})^{}_{\alpha J}
(M^\dagger_{\rm D} M^{}_{\rm D})^{}_{IJ}\right] {\cal F} \left( \frac{M^2_J}{M^2_I} \right) \right. \nonumber \\
&  &
+ \left. {\rm Im}\left[(M^*_{\rm D})^{}_{\alpha I} (M^{}_{\rm D})^{}_{\alpha J} (M^\dagger_{\rm D} M^{}_{\rm D})^*_{IJ}\right] {\cal G}  \left( \frac{M^2_J}{M^2_I} \right) \right\} \; ,
\label{3.4}
\end{eqnarray}
with ${\cal F}(x) = \sqrt{x} \{(2-x)/(1-x)+ (1+x) \ln [x/(1+x)] \}$ and ${\cal G}(x) = 1/(1-x)$. And $0<\kappa(\widetilde m^{}_I) <1$ is the efficiency factor, whose value is determined by the washout mass parameter
\begin{eqnarray}
\widetilde m^{}_I = \sum^{}_\alpha \widetilde m^{}_{I \alpha} = \sum^{}_\alpha  \frac{|(M^{}_{\rm D})^{}_{\alpha I}|^2}{M^{}_I} \; .
\label{3.5}
\end{eqnarray}
The dependence of $\kappa(\widetilde m)$ on $\widetilde m$ can be well described by the following analytical fits \cite{giudice}
\begin{eqnarray}
\frac{1}{\kappa(\widetilde m)} \simeq \frac{3.3 \times 10^{-3} ~{\rm eV}}{\widetilde m} + \left( \frac{\widetilde m} {5.5 \times 10^{-4} ~{\rm eV}} \right)^{1.16} \;.
\label{3.6}
\end{eqnarray}
In a realistic seesaw model, one usually has $\widetilde m \gtrsim \sqrt{\Delta m^2_{21}} \simeq 8.7 \times 10^{-3}$ eV (see below), in which regime $\kappa(\widetilde m)$ is roughly inversely proportional to $\widetilde m$.

For the scenario studied in the present paper, due to the $\mu$-$\tau$ reflection symmetry, one has $\varepsilon^{}_{I e} =0$ and $\varepsilon^{}_{I \mu} =- \varepsilon^{}_{I \tau}$. This leads $\varepsilon^{}_I$ and subsequently $Y^{}_{\rm B}$ to be vanishing. However, in the two-flavor regime which holds in the temperature range $10^{9}$---$10^{12}$ GeV \cite{flavor}, a successful leptogenesis may become possible. This is because in this regime the lepton asymmetries from $N^{}_I$ are stored along the $\ket{L^{}_\tau}$ and $\ket{L^{}_{I\gamma}}$ directions in the flavor space with
\begin{eqnarray}
\ket{L^{}_{I \gamma}} = \frac{1}{ \sqrt{\left|(M^{}_{\rm D})^{}_{e I}\right|^2 + \left|(M^{}_{\rm D})^{}_{\mu I}\right|^2} } \left[(M^{}_{\rm D})^*_{e I} \ket{L^{}_e} + (M^{}_{\rm D})^*_{\mu I} \ket{L^{}_\mu}\right] \;,
\label{3.7}
\end{eqnarray}
and they are subject to different washout effects:
\begin{eqnarray}
Y^{}_{\rm B}
=  -c r \left[ \varepsilon^{}_{I \gamma} \kappa \left(\frac{417}{589} \widetilde m^{}_{I \gamma} \right) + \varepsilon^{}_{I \tau} \kappa \left(\frac{390}{589} \widetilde m^{}_{I \tau} \right) \right]
 \;,
\label{3.8}
\end{eqnarray}
with $\varepsilon^{}_{I \gamma} = \varepsilon^{}_{I e} + \varepsilon^{}_{I \mu}$ and $\widetilde m^{}_{I \gamma} = \widetilde m^{}_{I e} + \widetilde m^{}_{I \mu}$.
Under the $\mu$-$\tau$ reflection symmetry, Eq.~(\ref{3.8}) simplifies to \cite{MN}
\begin{eqnarray}
Y^{}_{\rm B}
=  -c r \varepsilon^{}_{I \mu} \left[ \kappa \left(\frac{417}{589} \widetilde m^{}_{I \gamma} \right)  -\kappa \left(\frac{390}{589} \widetilde m^{}_{I \tau} \right) \right]
 \;.
\label{3.9}
\end{eqnarray}
We see that $Y^{}_{\rm B}$ will be non-vanishing unless $390 \widetilde m^{}_{I \tau}$ coincides with $417  \widetilde m^{}_{I \gamma}$. For this reason, we will work in the two-flavor regime in the following discussions.

\subsection{Implications of TM1 $\mu$-$\tau$ reflection symmetry for leptogenesis}

In this subsection, we study the implications of the TM1 $\mu$-$\tau$ reflection symmetry for leptogenesis. Generally speaking, the final baryon asymmetry is mainly owing to the lightest right-handed neutrino, because the lepton asymmetries from heavier right-handed neutrinos are likely to be erased by its lepton-number-violating interactions. However, because of $(M^\dagger_{\rm D} M^{}_{\rm D})^{}_{12} = (M^\dagger_{\rm D} M^{}_{\rm D})^{}_{13} =0$ (see Eq.~(\ref{2.1.5})), the CP asymmetries $\varepsilon^{}_{1\alpha}$ for the decay processes of $N^{}_1$ are vanishing. Therefore, the final baryon asymmetry can only originate from $N^{}_2$ or $N^{}_3$. But the lepton asymmetries from them may be subject to the washout effects from $N^{}_1$ when it is the lightest one.

Since $N^{}_2$ and $N^{}_3$ are on an equal footing (see Eq.~(\ref{2.1.5})), we choose to study the case of $M^{}_2< M^{}_3$ (where the final baryon asymmetry is mainly owing to $N^{}_2$), while the results for the case of $M^{}_3< M^{}_2$ (where the final baryon asymmetry is mainly owing to $N^{}_3$) are completely similar.
Let us first consider the case of $M^{}_2 < M^{}_1$, where the washout effects from $N^{}_1$ are irrelevant. The baryon asymmetry from $N^{}_2$ can be calculated according to Eq.~(\ref{3.9}) with
\begin{eqnarray}
&& \varepsilon^{}_{2 \mu} = \frac{M^{}_3}{8\pi v^2} \frac{(be-cd)(3bd+2ce)}{3b^2+2c^2} \left[ {\cal G}  \left( \frac{M^2_3}{M^2_2} \right) + \eta^{}_2 {\cal F}  \left( \frac{M^2_3}{M^2_2} \right) \right] \;, \nonumber \\
&& \widetilde m^{}_{2 \gamma} = 2 b^2 + c^2 \;, \hspace{1cm} \widetilde m^{}_{2 \tau} = b^2 + c^2 \;.
\label{3.1.1}
\end{eqnarray}
Taking account the Casas-Ibarra parametrization for $M^{}_{\rm D}$ (see Eqs.~(\ref{2.1.14}, \ref{2.1.16}, \ref{2.1.18})), the results in Eq.~(\ref{3.1.1}) are explicitly expressed as
\begin{eqnarray}
&& \varepsilon^{}_{2 \mu} = \frac{M^{}_3}{8\sqrt{6}\pi v^2} \frac{\sqrt{m^{}_2 m^{}_3 }(m^{}_3 - m^{}_2) \cos \vartheta \sin \vartheta}{m^{}_2 \cos^2 \vartheta + m^{}_3 \sin^2 \vartheta} \left[ {\cal G}  \left( \frac{M^2_3}{M^2_2} \right) +  {\cal F}  \left( \frac{M^2_3}{M^2_2} \right) \right] \;, \nonumber \\
&& \widetilde m^{}_{2 \gamma} = \frac{2}{3} m^{}_2 \left(1- \frac{3}{4} s^2_{13} \right) \cos^2 \vartheta  + \frac{1}{2} m^{}_3 \left(1 + s^2_{13} \right) \sin^2 \vartheta \mp \sqrt{m^{}_2 m^{}_3} s^{}_{12} c^{}_{13} s^{}_{13} \cos \vartheta \sin \vartheta  \;,\nonumber \\
&& \widetilde m^{}_{2 \tau} = \frac{1}{3} m^{}_2  \left(1 + \frac{3}{2} s^2_{13} \right) \cos^2 \vartheta + \frac{1}{2} m^{}_3 \left(1 - s^2_{13} \right) \sin^2 \vartheta  \pm \sqrt{m^{}_2 m^{}_3} s^{}_{12} c^{}_{13} s^{}_{13} \cos \vartheta \sin \vartheta \;,
\label{3.1.2}
\end{eqnarray}
for $\eta^{}_2 =1$, and
\begin{eqnarray}
&& \varepsilon^{}_{2 \mu} = \frac{M^{}_3}{8\sqrt{6}\pi v^2} \frac{\sqrt{m^{}_2 m^{}_3 }(m^{}_3 + m^{}_2) \cosh \vartheta \sinh \vartheta}{m^{}_2 \cosh^2 \vartheta + m^{}_3 \sinh^2 \vartheta} \left[ {\cal F}  \left( \frac{M^2_3}{M^2_2} \right) - {\cal G}  \left( \frac{M^2_3}{M^2_2} \right)  \right] \;, \nonumber \\
&& \widetilde m^{}_{2 \gamma} = \frac{2}{3} m^{}_2 \left(1- \frac{3}{4} s^2_{13} \right) \cosh^2 \vartheta  + \frac{1}{2} m^{}_3 \left(1 + s^2_{13} \right) \sinh^2 \vartheta \pm \sqrt{m^{}_2 m^{}_3} s^{}_{12} c^{}_{13} s^{}_{13} \cosh \vartheta \sinh \vartheta  \;,\nonumber \\
&& \widetilde m^{}_{2 \tau} = \frac{1}{3} m^{}_2  \left(1 + \frac{3}{2} s^2_{13} \right) \cosh^2 \vartheta + \frac{1}{2} m^{}_3 \left(1 - s^2_{13} \right) \sinh^2 \vartheta  \mp \sqrt{m^{}_2 m^{}_3} s^{}_{12} c^{}_{13} s^{}_{13} \cosh \vartheta \sinh \vartheta \;,
\label{3.1.3}
\end{eqnarray}
for $\eta^{}_2 =-1$. It is found that the expressions of $\varepsilon^{}_{2 \mu}$ are same for $\delta = -\pi/2$ and $\pi/2$. And, in the expressions of $\widetilde m^{}_{2 \gamma}$ and $\widetilde m^{}_{2 \tau}$, the terms that depend on the signs of $\delta$ are suppressed by $s^{}_{13}$ relative to the dominant terms. For this reason, we will only show the results for $\delta = -\pi/2$ as representatives in the following numerical calculations. We have numerically checked that the results for $\delta = \pi/2$ only show slight quantitative but no significant qualitative differences.
Furthermore, one has $\widetilde m^{}_{2 \gamma}> \widetilde m^{}_{2 \tau} \gtrsim \sqrt{\Delta m^2_{21}}$, for which $\kappa (417/589 \widetilde m^{}_{2 \gamma} )  -\kappa (390/589 \widetilde m^{}_{2 \tau} )$ is negative (see Eq.~(\ref{3.6})). In this case, in order to get a positive $Y^{}_{\rm B}$, $\varepsilon^{}_{2 \mu}$ needs to be positive.

\begin{figure*}[t]
\centering
\includegraphics[width=6.5in]{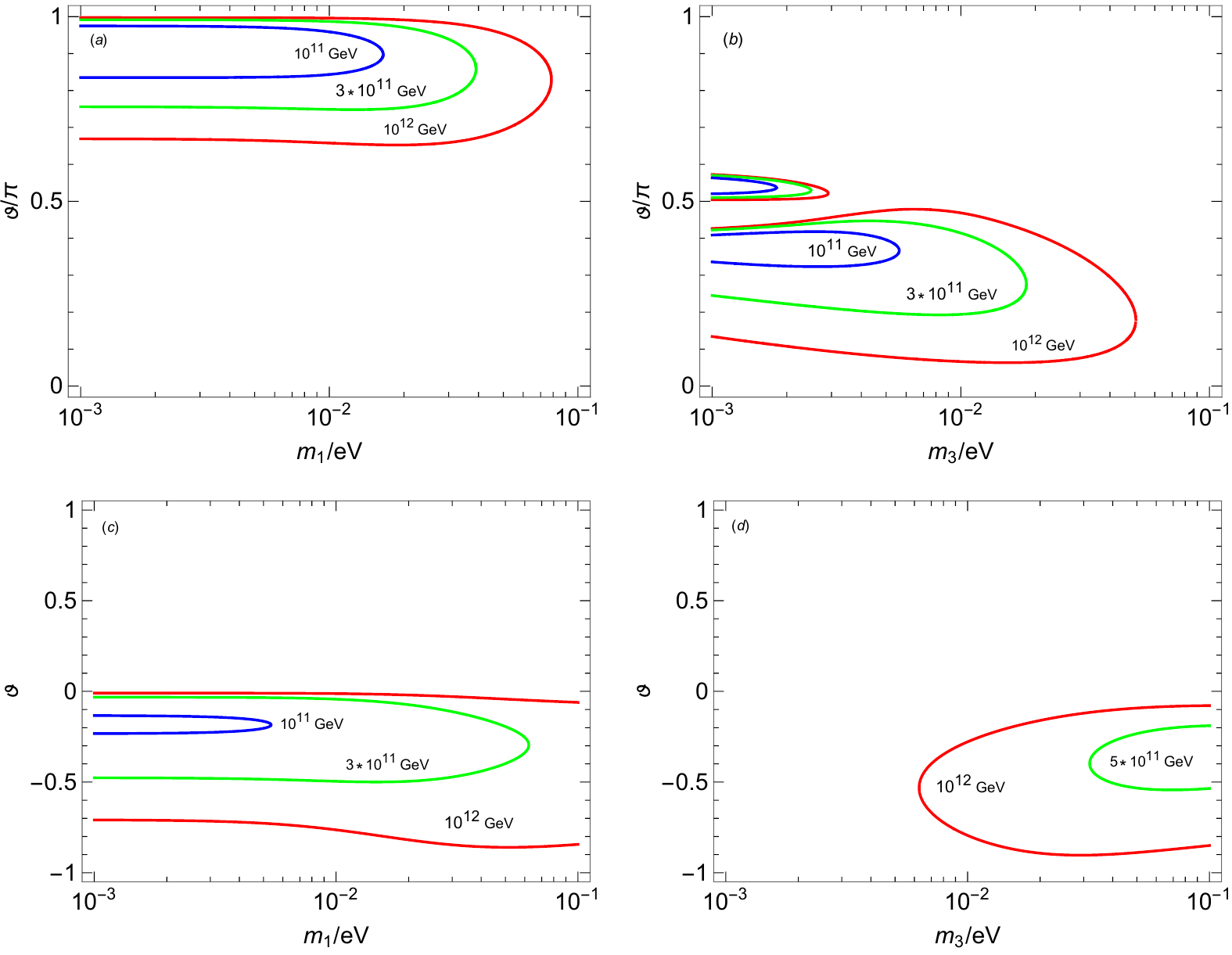}
\caption{ In the TM1 $\mu$-$\tau$ reflection symmetry scenario and the case of $M^{}_2 < M^{}_1$, for some benchmark values of $M^{}_2$ (see the labels for the plotted lines), the values of $\vartheta$ versus the lightest neutrino mass for leptogenesis to be viable: the NO case with $\eta^{}_2 =1$ (a); the IO case with $\eta^{}_2 =1$ (b); the NO case with $\eta^{}_2 =-1$ (c); the IO case with $\eta^{}_2 =-1$ (d).  }
\label{fig5}
\end{figure*}

\begin{figure*}[t]
\centering
\includegraphics[width=6.5in]{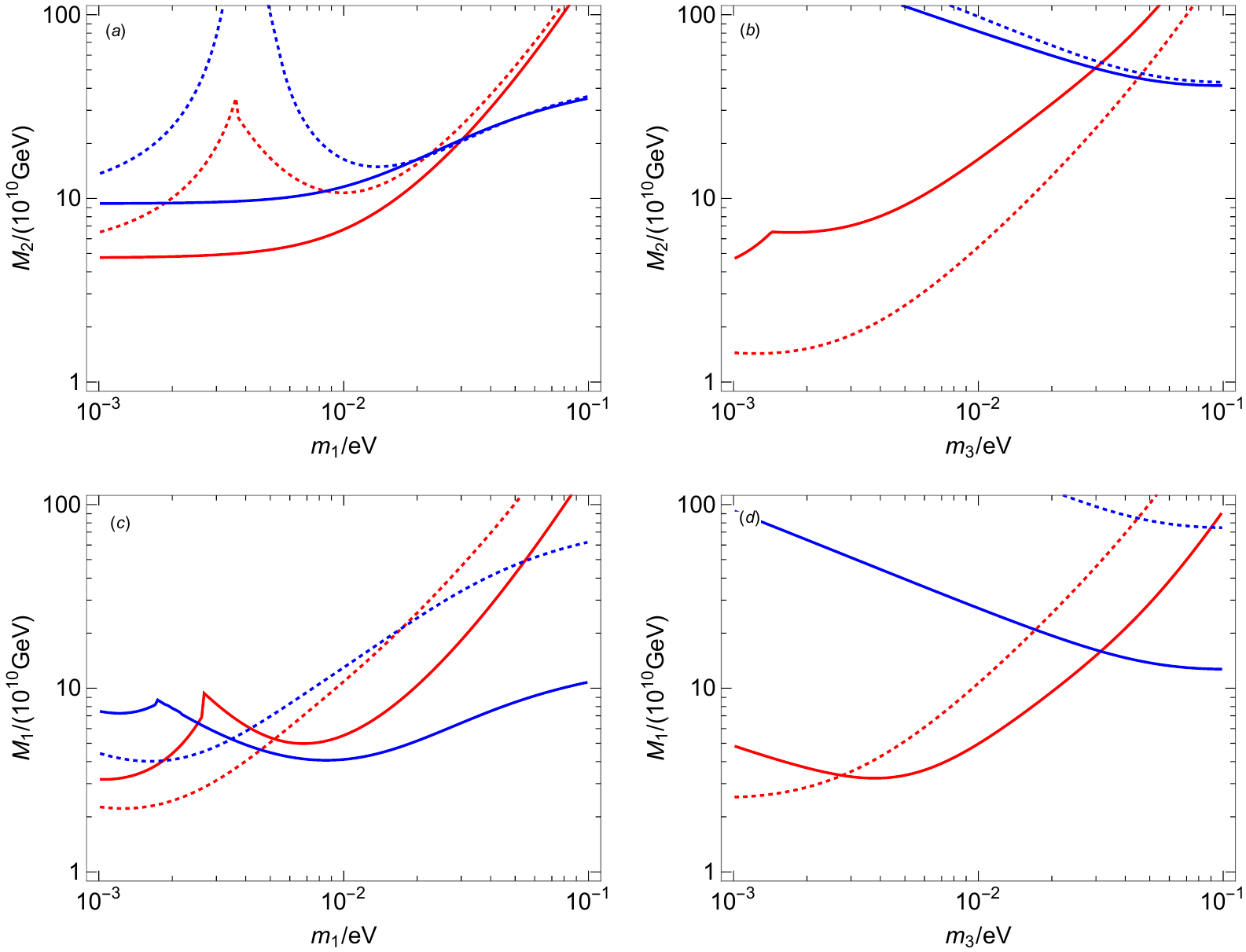}
\caption{ Upper: in the TM1 $\mu$-$\tau$ reflection symmetry scenario, the lower bound on $M^{}_2$ for leptogenesis to be viable as functions of the lightest neutrino mass in the NO (a) and IO (b) cases.
The red and blue solid (dotted) lines respectively correspond to $\eta^{}_2 = 1$ and $-1$ in the case of $M^{}_2 < M^{}_1$ ($M^{}_1 < M^{}_2$). Lower: in the TM2 $\mu$-$\tau$ reflection symmetry scenario, the lower bound on $M^{}_1$ for leptogenesis to be viable as functions of the lightest neutrino mass in the NO (c) and IO (d) cases.
The red and blue solid (dotted) lines respectively correspond to $\eta^{}_1 = 1$ and $-1$ in the case of $M^{}_1 < M^{}_2$ ($M^{}_2 < M^{}_1$).}
\label{fig6}
\end{figure*}

In Fig.~\ref{fig5}, for some benchmark values of $M^{}_2$, we plot the values of $\vartheta$ versus the lightest neutrino mass for leptogenesis to be viable in the NO and IO cases with $\eta^{}_2 =\pm 1$. In obtaining these results, $M^{}_3/M^{}_2 =3$ has been taken as a benchmark value. In fact, the dependence of the produced baryon asymmetry on the concrete value of $M^{}_3/M^{}_2$ is quite weak provided that $N^{}_2$ and $N^{}_3$ are not nearly degenerate. Since the two-flavor regime would become invalid beyond $10^{12}$ GeV, the parameter spaces of $\vartheta$ versus the lightest neutrino mass for leptogenesis to be viable are those enclosed by the lines corresponding to $M^{}_2 = 10^{12}$ GeV. For $\eta^{}_2 =1$, there is an upper bound about 0.1 eV on the lightest neutrino mass for leptogenesis to be viable. This is because, for a too large lightest neutrino mass, $\varepsilon^{}_{2 \mu}$ (due to the factor $m^{}_3-m^{}_2$) and the efficiency factors would get highly suppressed (see Eq.~(\ref{3.1.2})). In the NO (or IO) case, $\vartheta \sim \pi$ (or $\pi/2$) are advantageous to leptogenesis, because $\varepsilon^{}_{2 \mu}$ (due to the factor $m^{}_2 \cos^2 \vartheta + m^{}_3 \sin^2 \vartheta$) and the efficiency factors would get more suppressed for $\vartheta \sim \pi/2$ (or $\pi$). For $\eta^{}_2 =-1$, large values of the lightest neutrino mass can also accommodate viable leptogenesis, because now $\varepsilon^{}_{2 \mu}$ is proportional to $m^{}_3+m^{}_2$ (see Eq.~(\ref{3.1.3})). In particular, in the IO case, $m^{}_3$ needs to be larger than 0.06 eV in order for leptogenesis to be viable. On the other hand, small values of $\vartheta$ are advantageous to leptogenesis, because the washout mass parameters would increase rapidly for large values of $\vartheta$.
Finally, in Fig.~\ref{fig6}(a) and (b) (for the NO and IO cases, respectively), we plot the lower bound on $M^{}_2$ for leptogenesis to be viable as functions of the lightest neutrino mass, obtained by allowing $\vartheta$ to vary in the whole range. Here the red and blue solid lines correspond to the cases of $\eta^{}_2 = 1$ and $-1$, respectively.
We see that $M^{}_2$ needs to be larger than a few $10^{10}$ GeV at least in order to make leptogenesis viable. In particular, in the IO case with $\eta^{}_2 =-1$, the parameter space of $M^{}_2$ for leptogenesis to be viable is rather narrow.

\begin{figure*}[t]
\centering
\includegraphics[width=6.5in]{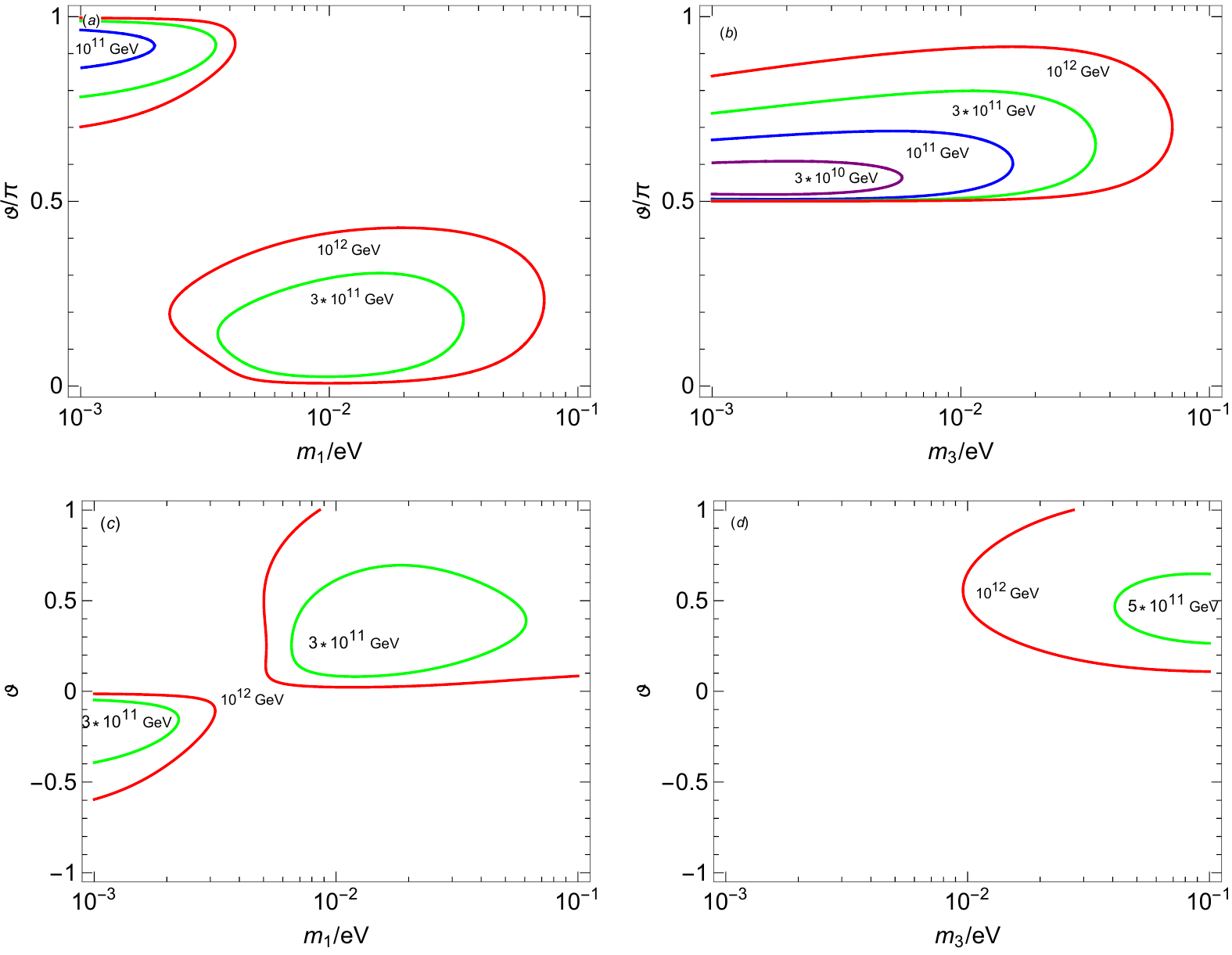}
\caption{ Same as Fig.~\ref{fig5}, expect that these results are for the case of $ M^{}_1 < M^{}_2$. }
\label{fig7}
\end{figure*}

Then, we consider the case of $ M^{}_1 < M^{}_2$. In this case, the lepton asymmetries from  $N^{}_2$ will be subject to the washout effects from $N^{}_1$, which are along the $\ket{L^{}_\tau}$ and $\ket{L^{}_{1\gamma}}$ directions.
The final baryon asymmetry can be calculated according to
\begin{eqnarray}
Y^{}_{\rm B}  = -c r  \varepsilon^{}_{2\mu}  \left\{ \kappa \left(\frac{417}{589} \widetilde m^{}_{2 \gamma} \right) \left[ \left(1-p^{1\gamma}_{2\gamma} \right)
+ p^{1\gamma}_{2\gamma} \exp\left(-\frac{3\pi \widetilde m^{}_{1\gamma}}{8 m^{}_*}\right) \right] -
 \kappa \left(\frac{390}{589} \widetilde m^{}_{2 \tau} \right) \exp\left(-\frac{3\pi \widetilde m^{}_{1\tau}}{8 m^{}_*}\right)  \right\} \;,
\label{3.1.4}
\end{eqnarray}
with $m^{}_* \simeq 1.1 \times 10^{-3}$ eV, $\widetilde m^{}_{1\tau} = a^2 = m^{}_1/6$ and $\widetilde m^{}_{1\gamma} = 5 a^2 = 5m^{}_1/6$. And
\begin{eqnarray}
p^{1\gamma}_{2\gamma} \equiv \left|\braket{ L^{}_{1\gamma} | L^{}_{2\gamma}}\right|^2 = \frac{ b^2 + c^2 }{ 5 (2 b^2 + c^2) } = \frac{\widetilde m^{}_{2 \tau}}{5 \widetilde m^{}_{2 \gamma}} \;,
\label{3.1.5}
\end{eqnarray}
measures the overlapping degree of the $\ket{L^{}_{1\gamma}}$ and $\ket{L^{}_{2\gamma}}$ directions.
It is obvious that $p^{1\gamma}_{2\gamma}$ falls in the range 1/10$-$1/5, implying that the washout effect along the $\ket{L^{}_{2\gamma}}$ direction is weak irrespective of the value of $\widetilde m^{}_{1\gamma}$. On the other hand, the strength of the washout effect along the $\ket{L^{}_{\tau}}$ direction can be either strong or weak depending on the value of $\widetilde m^{}_{1\tau}$.
In the $\widetilde m^{}_{1\tau} \ll m^{}_*$ regime, this effect is also weak. In this regime, the results of leptogenesis are basically same as in the case of $M^{}_2 < M^{}_1$. But in the $\widetilde m^{}_{1\tau} \gg m^{}_*$ regime, this effect is strong, suppressing the term proportional to $\kappa (390/589 \widetilde m^{}_{2 \tau})$ in Eq.~(\ref{3.1.4}). In this regime, in order to get a positive $Y^{}_{\rm B}$, $\varepsilon^{}_{2 \mu}$ needs to be negative instead. These observations can help us understand the results in Fig.~\ref{fig7} which, for some benchmark values of $M^{}_2$, plots the values of $\vartheta$ versus the lightest neutrino mass for leptogenesis to be viable in the NO and IO cases with $\eta^{}_2 =\pm 1$: in the NO case, the results are basically same as in the case of $M^{}_2 < M^{}_1$ for small values of $m^{}_1$ (corresponding to $\widetilde m^{}_{1\tau} \ll m^{}_*$), while $\cos \vartheta \sin \vartheta$ (or $\cosh \vartheta \sinh \vartheta$) takes an opposite sign against in the case of $M^{}_2 < M^{}_1$ for large values of $m^{}_1$ (corresponding to $\widetilde m^{}_{1\tau} \gg m^{}_*$); in the IO case, $\cos \vartheta \sin \vartheta$ (or $\cosh \vartheta \sinh \vartheta$) takes an opposite sign against in the case of $M^{}_2 < M^{}_1$ for the whole range of $m^{}_3$ (which always gives $\widetilde m^{}_{1\tau} \gg m^{}_*$). Finally, in Fig.~\ref{fig6}(a) and (b), we plot the lower bound on $M^{}_2$ for leptogenesis to be viable as functions of the lightest neutrino mass, where the red and blue dotted lines respectively correspond to the cases of $\eta^{}_2 = 1$ and $-1$. We see that in most cases the washout effects from $N^{}_1$ would more or less narrow the parameter space of $M^{}_2$ for leptogenesis to be viable. But in the IO case with $\eta^{}_2 = 1$, the reverse is true.

\subsection{Implications of TM2 $\mu$-$\tau$ reflection symmetry for leptogenesis}

In this subsection, we study the implications of the TM2 $\mu$-$\tau$ reflection symmetry for leptogenesis.
Now, because of $(M^\dagger_{\rm D} M^{}_{\rm D})^{}_{12} = (M^\dagger_{\rm D} M^{}_{\rm D})^{}_{23} =0$ (see Eq.~(\ref{2.2.3})), the CP asymmetries $\varepsilon^{}_{2\alpha}$ for the decay processes of $N^{}_2$ are vanishing. Therefore, the final baryon asymmetry can only originate from $N^{}_1$ or $N^{}_3$. But the lepton asymmetries from them may be subject to the washout effects from $N^{}_2$ when it is the lightest one.

Since $N^{}_1$ and $N^{}_3$ are on an equal footing (see Eq.~(\ref{2.2.3})), we choose to study the case of $M^{}_1< M^{}_3$ (where the final baryon asymmetry is mainly owing to $N^{}_1$), while the results for the case of $M^{}_3< M^{}_1$ (where the final baryon asymmetry is mainly owing to $N^{}_3$) are completely similar.
Let us first consider the case of $M^{}_1 < M^{}_2$, where the washout effects from $N^{}_2$ are irrelevant. The baryon asymmetry from $N^{}_1$ can be calculated according to Eq.~(\ref{3.8}) with
\begin{eqnarray}
&& \varepsilon^{}_{1 \mu} = \frac{M^{}_3}{8\pi v^2} \frac{(cd-be)(3bd+ce)}{3b^2+c^2} \left[ {\cal G}  \left( \frac{M^2_3}{M^2_1} \right) + \eta^{}_1 {\cal F}  \left( \frac{M^2_3}{M^2_1} \right) \right] \;, \nonumber \\
&& \widetilde m^{}_{1 \gamma} = 5 b^2 + c^2 \;, \hspace{1cm} \widetilde m^{}_{1 \tau} = b^2 + c^2 \;.
\label{3.2.1}
\end{eqnarray}
Taking account the Casas-Ibarra parametrization for $M^{}_{\rm D}$ (see Eqs.~(\ref{2.1.14}, \ref{2.2.11}, \ref{2.2.12})), the results in Eq.~(\ref{3.2.1}) are explicitly expressed as
\begin{eqnarray}
&& \varepsilon^{}_{1 \mu} = - \frac{M^{}_3}{16\sqrt{3}\pi v^2} \frac{\sqrt{m^{}_1 m^{}_3 }(m^{}_3 - m^{}_1) \cos \vartheta \sin \vartheta}{m^{}_1 \cos^2 \vartheta + m^{}_3 \sin^2 \vartheta} \left[ {\cal G}  \left( \frac{M^2_3}{M^2_1} \right) +  {\cal F}  \left( \frac{M^2_3}{M^2_1} \right) \right] \;, \nonumber \\
&& \widetilde m^{}_{1 \gamma} = \frac{5}{6} m^{}_1 \left(1- \frac{3}{5} s^2_{13} \right) \cos^2 \vartheta  + \frac{1}{2} m^{}_3 \left(1 + s^2_{13} \right) \sin^2 \vartheta \mp \sqrt{m^{}_1 m^{}_3} c^{}_{12} c^{}_{13} s^{}_{13} \cos \vartheta \sin \vartheta  \;,\nonumber \\
&& \widetilde m^{}_{1 \tau} = \frac{1}{6} m^{}_1 \left(1 + 3 s^2_{13} \right) \cos^2 \vartheta  + \frac{1}{2} m^{}_3 \left(1 - s^2_{13} \right) \sin^2 \vartheta \pm \sqrt{m^{}_1 m^{}_3} c^{}_{12} c^{}_{13} s^{}_{13} \cos \vartheta \sin \vartheta  \;,
\label{3.2.2}
\end{eqnarray}
for $\eta^{}_1 =1$, and
\begin{eqnarray}
&& \varepsilon^{}_{1 \mu} = \frac{M^{}_3}{16\sqrt{3}\pi v^2} \frac{\sqrt{m^{}_1 m^{}_3 }(m^{}_3 + m^{}_1) \cosh \vartheta \sinh \vartheta}{m^{}_1 \cosh^2 \vartheta + m^{}_3 \sinh^2 \vartheta} \left[ {\cal G}  \left( \frac{M^2_3}{M^2_1} \right)- {\cal F}  \left( \frac{M^2_3}{M^2_1} \right) \right] \;, \nonumber \\
&& \widetilde m^{}_{1 \gamma} = \frac{5}{6} m^{}_1 \left(1- \frac{3}{5} s^2_{13} \right) \cosh^2 \vartheta  + \frac{1}{2} m^{}_3 \left(1 + s^2_{13} \right) \sinh^2 \vartheta \pm \sqrt{m^{}_1 m^{}_3} c^{}_{12} c^{}_{13} s^{}_{13} \cosh \vartheta \sinh \vartheta  \;,\nonumber \\
&& \widetilde m^{}_{1 \tau} = \frac{1}{6} m^{}_1 \left(1 + 3 s^2_{13} \right) \cosh^2 \vartheta  + \frac{1}{2} m^{}_3 \left(1 - s^2_{13} \right) \sinh^2 \vartheta \mp \sqrt{m^{}_1 m^{}_3} c^{}_{12} c^{}_{13} s^{}_{13} \cosh \vartheta \sinh \vartheta  \;,
\label{3.2.3}
\end{eqnarray}
for $\eta^{}_1 =-1$. For a same reason as explained below Eq.~(\ref{3.1.3}), we will only show the results for $\delta = -\pi/2$ as representatives in the following numerical calculations. It is useful to note that the relation $\widetilde m^{}_{1 \gamma}> \widetilde m^{}_{1 \tau} \gtrsim \sqrt{\Delta m^2_{21}}$ holds, for which $\kappa (417/589 \widetilde m^{}_{1 \gamma} )  -\kappa (390/589 \widetilde m^{}_{1 \tau} )$ is negative (see Eq.~(\ref{3.6})). In this case, in order to get a positive $Y^{}_{\rm B}$, $\varepsilon^{}_{1 \mu}$ needs to be positive.

\begin{figure*}[t]
\centering
\includegraphics[width=6.5in]{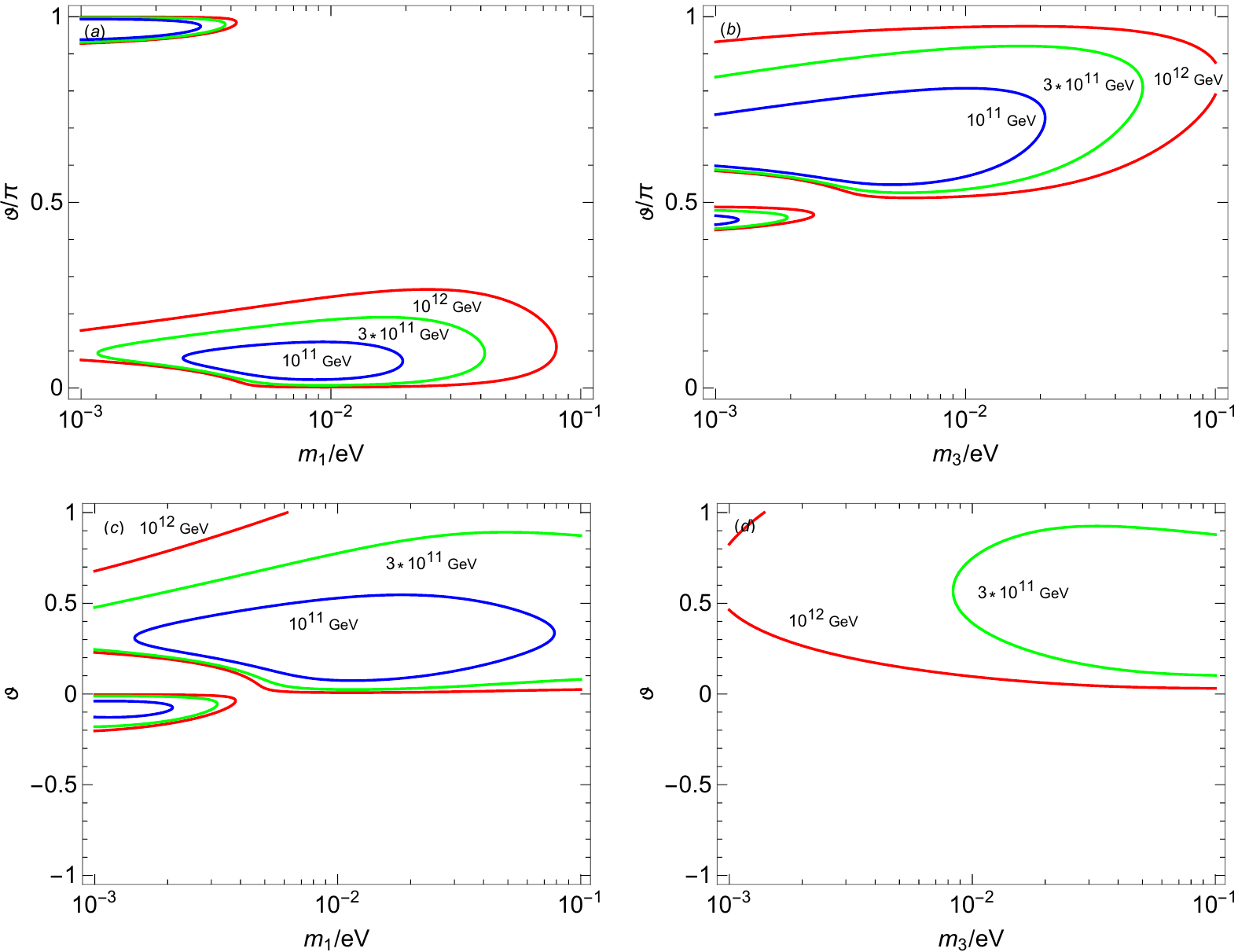}
\caption{ In the TM2 $\mu$-$\tau$ reflection symmetry scenario and the case of $M^{}_1 < M^{}_2$, for some benchmark values of $M^{}_1$ (see the labels for the plotted lines), the values of $\vartheta$ versus the lightest neutrino mass for leptogenesis to be viable: the NO case with $\eta^{}_1 =1$ (a); the IO case with $\eta^{}_1 =1$ (b); the NO case with $\eta^{}_1 =-1$ (a); the IO case with $\eta^{}_1 =-1$ (d). }
\label{fig8}
\end{figure*}

In Fig.~\ref{fig8}, for some benchmark values of $M^{}_1$, we plot the values of $\vartheta$ versus the lightest neutrino mass for leptogenesis to be viable in the NO and IO cases with $\eta^{}_1 =\pm 1$. In obtaining these results, $M^{}_3/M^{}_1 =3$ has been taken as a benchmark value. The results here have a lot in common with those in last subsection (see there for more detailed discussions): for $\eta^{}_1 =1$, there is an upper bound about 0.1 eV on the lightest neutrino mass for leptogenesis to be viable. In the NO (or IO) case, $\vartheta \sim 0$ (or $\pi/2$) are advantageous to leptogenesis. For $\eta^{}_1 =-1$, large values of the lightest neutrino mass can also accommodate viable leptogenesis. And small values of $\vartheta$ are advantageous to leptogenesis.
Finally, in Fig.~\ref{fig6}(c) and (d) (for the NO and IO cases, respectively), we plot the lower bound on $M^{}_1$ for leptogenesis to be viable as functions of the lightest neutrino mass, where the red and blue solid lines respectively correspond to the cases of $\eta^{}_1 = 1$ and $-1$.
We see that $M^{}_1$ needs to be larger than a few $10^{10}$ GeV at least in order to make leptogenesis viable.

\begin{figure*}[t]
\centering
\includegraphics[width=6.5in]{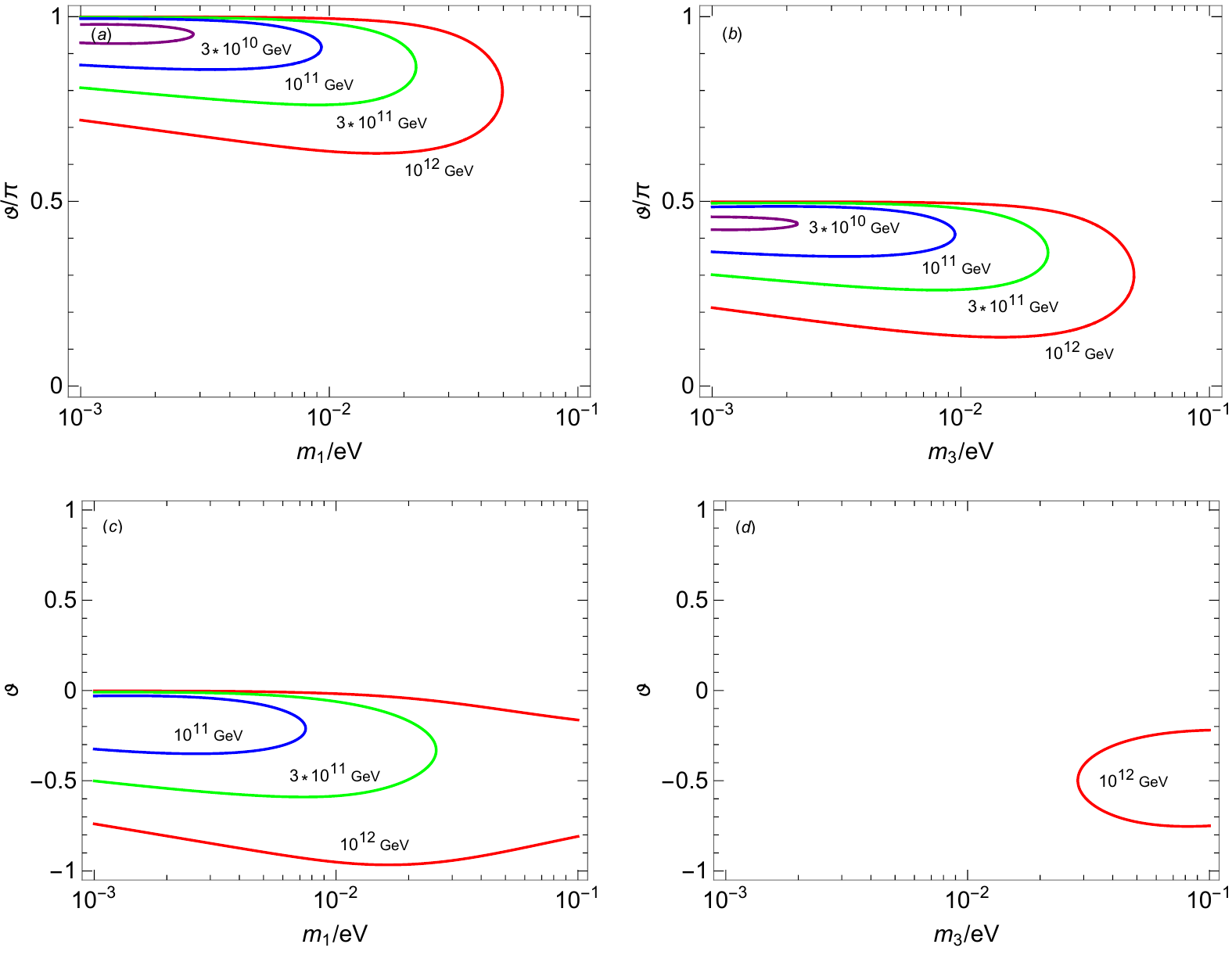}
\caption{ Same as Fig.~8, except that these results are for the case of $ M^{}_2 < M^{}_1$. }
\label{fig9}
\end{figure*}

Then, we consider the case of $ M^{}_2 < M^{}_1$. In this case, the lepton asymmetries from  $N^{}_1$ will be subject to the washout effects from $N^{}_2$, which are along the $\ket{L^{}_\tau}$ and $\ket{L^{}_{2\gamma}}$ directions.
The final baryon asymmetry can be calculated according to
\begin{eqnarray}
Y^{}_{\rm B}  = - c r  \varepsilon^{}_{1\mu}  \left\{ \kappa \left(\frac{417}{589} \widetilde m^{}_{1 \gamma} \right) \left[ \left(1-p^{2\gamma}_{1\gamma} \right)
+ p^{2\gamma}_{1\gamma} \exp\left(-\frac{3\pi \widetilde m^{}_{2\gamma}}{8 m^{}_*}\right) \right] - \kappa \left(\frac{390}{589} \widetilde m^{}_{1 \tau} \right) \exp\left(-\frac{3\pi \widetilde m^{}_{2\tau}}{8 m^{}_*}\right)   \right\} \;,
\label{3.2.4}
\end{eqnarray}
with $\widetilde m^{}_{2\tau} = a^2 = m^{}_2/3$, $\widetilde m^{}_{2\gamma} = 2 a^2 = 2 m^{}_2/3$ and
\begin{eqnarray}
p^{2\gamma}_{1\gamma} \equiv \left|\braket{ L^{}_{2\gamma} | L^{}_{1\gamma}}\right|^2 = \frac{ b^2 + c^2 }{ 2 (5 b^2 + c^2) } = \frac{\widetilde m^{}_{1 \tau}}{2 \widetilde m^{}_{1 \gamma}} \;.
\label{3.2.5}
\end{eqnarray}
It is obvious that $p^{1\gamma}_{2\gamma}$ falls in the range 1/10$-$1/2, implying that the washout effect along the $\ket{L^{}_{1\gamma}}$ direction is more likely to be weak. On the other hand, due to $\widetilde m^{}_{2\tau} = m^{}_2/3$, the washout effect along the $\ket{L^{}_{\tau}}$ direction is strong, suppressing the term proportional to $\kappa (390/589 \widetilde m^{}_{2 \tau})$ in Eq.~(\ref{3.2.4}). In this regime, in order to get a positive $Y^{}_{\rm B}$, $\varepsilon^{}_{1 \mu}$ needs to be negative instead. This observation can help us understand the results in Fig.~\ref{fig9} which, for some benchmark values of $M^{}_1$, plots the values of $\vartheta$ versus the lightest neutrino mass for leptogenesis to be viable in the NO and IO cases with $\eta^{}_1 =\pm 1$: $\cos \vartheta \sin \vartheta$ (or $\cosh \vartheta \sinh \vartheta$) takes an opposite sign against in the case of $M^{}_1 < M^{}_2$. Finally, in Fig.~\ref{fig6}(c) and (d), we plot the lower bound on $M^{}_1$ for leptogenesis to be viable as functions of the lightest neutrino mass, where the red and blue dotted lines respectively correspond to the cases of $\eta^{}_1 = 1$ and $-1$. We see that the washout effects from $N^{}_2$ would more or less narrow the parameter space of $M^{}_1$ for leptogenesis to be viable.

\section{Trimaximal $\mu$-$\tau$ reflection symmetry in the minimal seesaw model}

In this section, we consider the combination of the trimaximal mixings and $\mu$-$\tau$ reflection symmetry in the more restrictive and predictive minimal seesaw model with only two right-handed neutrinos \cite{minimal}.

Let us first consider the combination of the TM1 mixing and $\mu$-$\tau$ reflection symmetry in the minimal seesaw model. This can be achieved by simply decoupling one right-handed neutrino from the seesaw model given in section~2.1 that accommodates the TM1 $\mu$-$\tau$ reflection symmetry. There are the following two qualitatively different choices.
The first one is to decouple $N^{}_1$ from $M^{}_{\rm D}$ in Eq.~(\ref{2.1.5}). The resulting $M^{}_\nu$ can be obtained from Eq.~(\ref{2.1.6}) by taking $a=0$. It still can be diagonalized by $U^{}_0$ in Eq.~(\ref{2.1.8}), giving the neutrino masses in Eq.~(\ref{2.1.10}). But now we have
$m^{\prime}_1 =0$, corresponding to the NO case. In this case, the fitted values of $T^{}_1$, $T^{}_2$ and $T^{}_3$ can be inferred from Fig.~\ref{fig1}(a) and (b) in the $m^{}_1 \to 0$ limit. And there is still one degree of freedom in reconstructing the model parameters from the measurable neutrino parameters. On the other hand, the predictions for the lepton flavor mixing parameters are same as those obtained in section~2.1, except that now $\rho$ becomes meaningless. And the results for the magnitudes of the Majorana neutrino mass matrix elements can be inferred from Fig.~\ref{fig3} in the $m^{}_1 \to 0$ limit.

The second choice is to decouple $N^{}_2$ (or equivalently $N^{}_3$) from $M^{}_{\rm D}$ in Eq.~(\ref{2.1.5}). The resulting $M^{}_\nu$ can be obtained from Eq.~(\ref{2.1.6}) by taking $b=c=0$. It still can be diagonalized by $U^{}_0$ in Eq.~(\ref{2.1.8}), giving the neutrino masses in Eq.~(\ref{2.1.10}). But now we have
$m^{\prime}_3 =0$, corresponding to the IO case. In this case, all the three model parameters (i.e., $a$, $d$ and $e$) can be completely determined from the measurable neutrino parameters. Their fitted values can be inferred from Fig.~\ref{fig1}(c) and (d) in the $m^{}_3 \to 0$ limit. On the other hand, the predictions for the lepton flavor mixing parameters are same as those obtained in section~2.1, except that now only the difference of $\rho$ and $\sigma$ is of physical meaning. And the results for the magnitudes of the Majorana neutrino mass matrix elements can be inferred from Fig.~\ref{fig4} in the $m^{}_3 \to 0$ limit.

Then, we consider the combination of the TM2 mixing and $\mu$-$\tau$ reflection symmetry in the minimal seesaw model. This can be achieved by simply decoupling one right-handed neutrino from the seesaw model given in section~2.2 that accommodates the TM2 $\mu$-$\tau$ reflection symmetry. Note that $N^{}_2$ can not serve for this purpose, because otherwise one would arrive at the unrealistic result $m^{\prime}_2 =0$ (see Eq.~(\ref{2.2.7})). Now we decouple $N^{}_1$ (or equivalently $N^{}_3$) from $M^{}_{\rm D}$ in Eq.~(\ref{2.2.3}). The resulting $M^{}_\nu$ can be obtained from Eq.~(\ref{2.2.4}) by taking $b=c=0$. It still can be diagonalized by $U^{}_0$ in Eq.~(\ref{2.2.5}), giving the neutrino masses in Eq.~(\ref{2.2.7}). But now we have $m^{\prime}_1 =0$ or $m^\prime_3 =0$, corresponding to the NO or IO case respectively. In this case, all the three model parameters (i.e., $a$, $d$ and $e$) can be completely determined from the measurable neutrino parameters. Their fitted values can be inferred from Fig.~\ref{fig2} in the $m^{}_1 \to 0$ or $m^{}_3 \to 0$ limit. On the other hand, the predictions for the lepton flavor mixing parameters are same as those obtained in section~2.2, except that now only the difference of $\rho$ and $\sigma$ is of physical meaning (or $\rho$ becomes meaningless) for the $m^{\prime}_1 =0$ (or $m^\prime_3 =0$) case. And, as mentioned in section~2.3, the results for the magnitudes of the Majorana neutrino mass matrix elements are very close to those in the TM1 $\mu$-$\tau$ reflection symmetry scenario.

\begin{figure*}[t]
\centering
\includegraphics[width=6.5in]{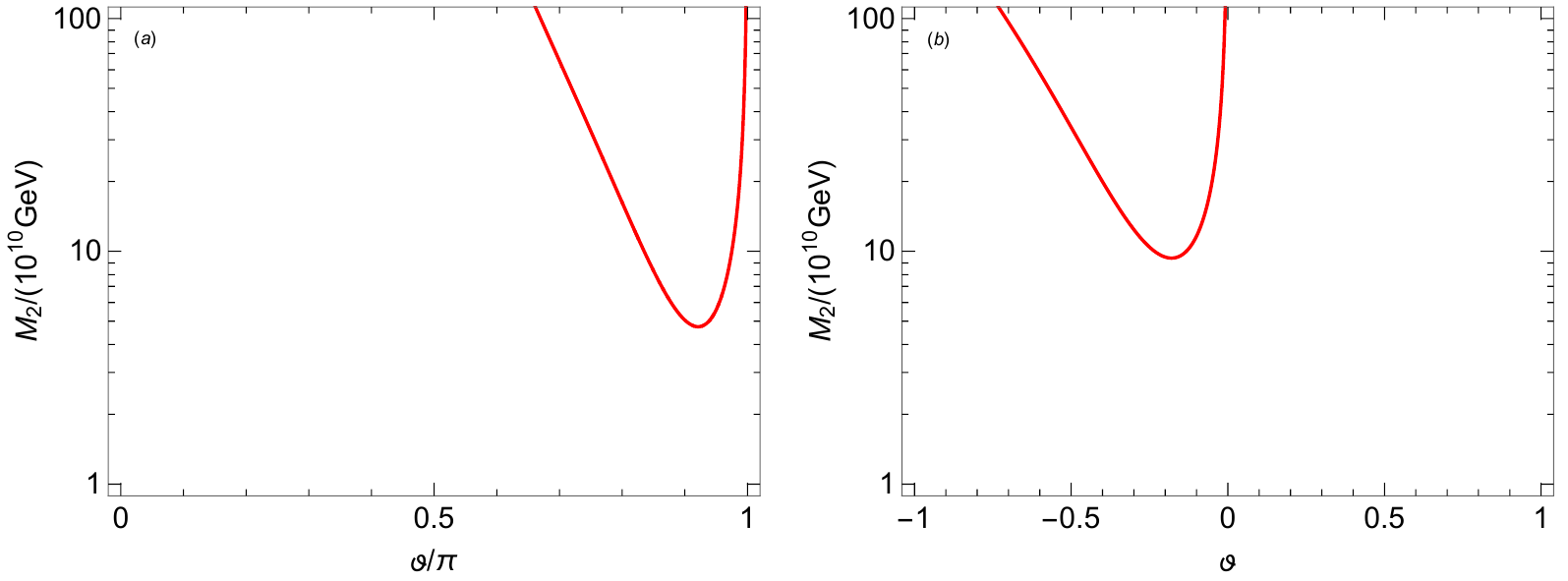}
\caption{ In the minimal seesaw model, for the TM1 $\mu$-$\tau$ reflection symmetry scenario and the case of $M^{}_2 < M^{}_3$, the values of $M^{}_2$ versus $\vartheta$ for leptogenesis to be viable in the NO case with $\eta^{}_2 =1$ (a) and $-1$ (b).}
\label{fig10}
\end{figure*}

Finally, we consider the implications of the trimaximal $\mu$-$\tau$ reflection symmetry in the minimal seesaw model for leptogenesis. It is easy to see that for the TM1 $\mu$-$\tau$ reflection symmetry scenario only in the NO case can leptogenesis have chance to be viable, while in the IO case leptogenesis can not proceed due to the orthogonality between two columns of $M^{}_{\rm D}$. Similarly, for the TM2 $\mu$-$\tau$ reflection symmetry scenario leptogenesis can not proceed neither in the NO case nor in the IO case. In Fig.~\ref{fig10}, for the TM1 $\mu$-$\tau$ reflection symmetry scenario and the case of $M^{}_2 < M^{}_3$, we plot the values of $M^{}_2$ versus $\vartheta$ for leptogenesis to be viable in the NO case with $\eta^{}_2 =\pm 1$. These results can be understood with the help of the discussions in section~3.1 in the $m^{}_1 \to 0$ limit.

\section{Summary}

As we know, the most natural way of generating the small neutrino masses is via the seesaw mechanism, which also provides an attractive explanation for the baryon asymmetry of the Universe via the leptogenesis mechanism. But the general type-I seesaw model with three right-handed neutrinos has the shortcoming that the number of its parameters is much larger than that of the measurable neutrino parameters, making itself lack predictive power.
In the literature, a typical approach to reducing its parameters and thus improving its predictability is to constrain its flavor structure by employing some flavor symmetry \cite{FS} such as the trimaximal symmetries and $\mu$-$\tau$ reflection symmetry, which are well motivated (both experimentally and theoretically) and
have interesting phenomenological consequences.

In this paper, we have made an attempt to combine the trimaximal (TM1 and TM2) mixings and $\mu$-$\tau$ reflection symmetry in the type-I seesaw model. Such a scenario is highly restrictive and predictive: there are only five model parameters in addition to three right-handed neutrino masses (see Eqs.~(\ref{2.1.5}, \ref{2.2.3})); all the lepton flavor mixing parameters except for $\theta^{}_{13}$ are predicted, leading to some definite predictions for the magnitudes of the Majorana neutrino mass matrix elements. We have derived the relations between the model parameters and the measurable neutrino parameters. It is found that we are left with one degree of freedom (i.e., the parameter $\vartheta$ here) in reconstructing the former from the latter. With the help of this degree of freedom, one can further reduce the model parameters by transforming one of them to vanishing. To facilitate the study on leptogenesis, we have further identified this degree of freedom as one parameter of the Casas-Ibarra parametrization for $M^{}_{\rm D}$.

We have also discussed a possible approach to get the desired mass matrices and studied the compatibility of the trimaximal $\mu$-$\tau$ reflection symmetry with texture zeros of $M^{}_\nu$. It is found that in the NO case $|M^{}_{ee}|$ may become vanishing at $m^{}_1 \simeq 2.5$ meV (or 6.8 meV) for $(\rho, \sigma) = (\pi/2, 0)$ (or $(0, \pi/2)$), while in the IO case $|M^{}_{\mu\tau}|$ may become vanishing at $m^{}_3 \simeq 0.02$ eV for $(\rho, \sigma) = (\pi/2, 0)$.

We have then studied the implications of the trimaximal $\mu$-$\tau$ reflection symmetry for leptogenesis. The results for the TM1 $\mu$-$\tau$ reflection symmetry scenario are summarized as follows, while those for the TM2 counterpart are similar. Due to the $\mu$-$\tau$ reflection symmetry, only in the two-flavor regime (which holds in the temperature range $10^{9} - 10^{12}$ GeV) can leptogenesis have chance to be viable. Due to the TM1 symmetry, the right-handed neutrino $N^{}_1$ can not be responsible for leptogenesis.
Considering that $N^{}_2$ and $N^{}_3$ are on an equal footing (see Eq.~(\ref{2.1.5})), we have chosen to study the case of $M^{}_2< M^{}_3$ (where the final baryon asymmetry is mainly owing to $N^{}_2$), while the results for the case of $M^{}_3< M^{}_2$ are completely similar. We have shown the parameter spaces of $\vartheta$ versus the lightest neutrino mass for leptogenesis to be viable in the NO and IO cases with $\eta^{}_2 = \pm 1$. For $\eta^{}_2 =1$, there is an upper bound about 0.1 eV on the lightest neutrino mass for leptogenesis to be viable. However, for $\eta^{}_2 =-1$, large values of the lightest neutrino mass can also accommodate viable leptogenesis. In particular, in the IO case, $m^{}_3$ needs to be larger than 0.06 eV in order for leptogenesis to be viable. We have also shown the lower bound on $M^{}_2$ for leptogenesis to be viable as functions of the lightest neutrino mass. The results show that $M^{}_2$ needs to be larger than a few $10^{10}$ GeV at least in order to make leptogenesis viable. In particular, in the IO case with $\eta^{}_2 =-1$, the parameter space of $M^{}_2$ for leptogenesis to be viable is rather narrow.
In the case of $N^{}_1$ being the lightest one, its washout effects may affect the lepton asymmetries from $N^{}_2$. It is found that the washout effect along the $\ket{L^{}_{2\gamma}}$ direction is weak, while the washout effect along the $\ket{L^{}_{\tau}}$ direction can be either weak or strong depending on the neutrino mass spectrum. In most cases (but in the IO case with $\eta^{}_2 = 1$) the washout effects from $N^{}_1$ would more or less narrow (expand instead) the parameter space of $M^{}_2$ for leptogenesis to be viable.

Furthermore, we have considered the combination of the trimaximal mixings and $\mu$-$\tau$ reflection symmetry in the minimal seesaw model. This can be achieved by simply decoupling one right-handed neutrino from the seesaw models given in section~2 that accommodate the trimaximal $\mu$-$\tau$ reflection symmetry. But it should be noted that, for the TM2 $\mu$-$\tau$ reflection symmetry scenario, a decoupling of $N^{}_2$ from $M^{}_{\rm D}$ in Eq.~(\ref{2.2.3}) would not give a realistic neutrino mass spectrum. As for leptogenesis, only in the TM1 $\mu$-$\tau$ reflection symmetry scenario and the NO case can it have chance to be viable.

Finally, we mention that, during the renormalization group evolution from the seesaw scale down to the electroweak scale, the differences among the Yukawa couplings $y^{}_\alpha$ of three charged leptons can induce the breaking of the $\mu$-$\tau$ reflection symmetry, making the lepton flavor mixing parameters deviate from the values given in Eq.~(\ref{5}) \cite{MTRrge}. However, in the Standard Model framework, due to the smallness of $y^{}_\alpha$, such effects are negligibly small. In comparison, in the Supersymmetric Standard Model framework, $y^{}_\alpha$ can be enhanced by a large $\tan{\beta}$ value so that the renormalization group evolution effects have chance to become sizable. The detailed studies show that, for large $\tan \beta$ values (e.g., $\tan \beta \gtrsim 30$), $\theta^{}_{23}$ has chance to deviate from $45^\circ$ by a few degrees while $\delta$ from $\pm 90^\circ$ by a few tens of degrees \cite{MTRrge}.

\vspace{0.5cm}

\underline{Acknowledgments} \vspace{0.2cm}

This work is supported in part by the National Natural Science Foundation of China under grant Nos.~11605081, 12142507 and 12147214, and the Natural Science Foundation of the Liaoning Scientific Committee under grant NO.~2019-ZD-0473.

\end{document}